\documentclass[aps, prd, twocolumn, preprintnumbers, nofootinbib, superscriptaddress, balancelastpage, 10pt]{revtex4-1}

\usepackage{graphicx}
\usepackage{amsmath, amsfonts, amssymb}
\usepackage{enumitem}
\usepackage{hyperref}
\usepackage{xcolor}
\usepackage{ulem}
\usepackage{makecell}
\usepackage{multirow}

\begin{document}

\title{Feasibility of ultrarelativistic bubbles in SMEFT}

\author{Upalaparna Banerjee}
\email{ubanerjee@uni-mainz.de}
\address{Department of Physics, Indian Institute of Technology, Kanpur 208016, India}
\address{PRISMA+ Cluster of Excellence \& Institute of Physics (THEP) \& Mainz Institute for Theoretical Physics, Johannes Gutenberg University, D-55099 Mainz, Germany}

\author{Sabyasachi Chakraborty}
\email{sabyac@iitk.ac.in}
\address{Department of Physics, Indian Institute of Technology, Kanpur 208016, India}

\author{Suraj Prakash}
\email{suraj.prakash@ific.uv.es}
\address{Department of Physics, Indian Institute of Technology, Kanpur 208016, India}
\address{Departament de Física Teòrica, IFIC (Universitat de València - CSIC), Parc Científic UV, C/ Catedrático José Beltrán 2, E-46980 Paterna (Valencia), Spain}

\author{Shakeel Ur Rahaman}
\email{shakeel.u.rahaman@durham.ac.uk}
\address{Institute for Particle Physics Phenomenology, Department of Physics, Durham University, Durham, DH1 3LE, United Kingdom}

\begin{abstract}
A first order electroweak phase transition probes physics beyond the Standard Model on multiple frontiers and therefore is of immense interest for theoretical exploration. We conduct a model-independent study of the effects of relevant dimension 6 and dimension 8 operators, of the Standard Model effective field theory, on electroweak phase transition. We use a thermally corrected and renormalization group improved potential and study its impact on nucleation temperature. We then outline bubble dynamics that lead to ultrarelativistic bubble wall velocities which are mainly motivated from the viewpoint of gravitational wave detection. We highlight the ranges of the Wilson coefficients that give rise to such bubble wall velocities and predict gravitational wave spectra generated by such transitions which can be tested in future experiments. 
\end{abstract}

\maketitle

\section{Introduction}\label{sec:introduction}
Gauge theories generically rely on phase transitions to generate masses for particles via spontaneous symmetry breaking.  For first order phase transition i.e., below a critical temperature the Universe undergoes a transition from a metastable state to a stable equilibrium state through the process of bubble nucleation, growth, and eventual merger. Physics pertaining to first order phase transition (FOPT) has received a lot of attention in the recent past and the reason is mainly twofold. Firstly, an FOPT implies a departure from thermal equilibrium which is a necessary criterion~\footnote{See for alternative scenarios using spontaneous baryogenesis~\cite{Cohen:1988kt}.} for explaining matter-antimatter asymmetry in the Universe \cite{Sakharov:1967dj,Kuzmin:1985mm}. This remains one of the most elusive shortcomings of the Standard Model (SM) of particle physics. In fact, the SM, with its specific set of parameters, measured with great precision by particle collision experiments, can only accommodate an adiabatic crossover transition at the weak scale~\cite{Aoki:1999fi,Csikor:1998eu,Laine:1998jb,Gurtler:1997hr}. Therefore, a first order electroweak phase transition (EWPT) is a natural testing ground for physics beyond the Standard Model (BSM). Secondly, cosmological FOPTs lead to the production of a gravitational wave spectrum (GWS)~\cite{Grojean:2006bp,Huang:2016cjm,Hashino:2016xoj,Beniwal:2017eik,Chala:2016ykx,Athron:2023xlk,Hashino:2022ghd} that can be detected by current and upcoming interferometric experiments. Thus, FOPT furnishes a complementary means to test BSM physics at colliders as well as at cosmic frontiers. An FOPT at the electroweak scale can lead to a GWS that falls within the frequency ranges of the Laser Interferometer Space Antenna (LISA)~\cite{LISA:2017pwj,Caprini:2015zlo,Caprini:2019egz}, Deci-hertz Interferometer Gravitational Wave Observatory (DECIGO)~\cite{Seto:2001qf} and Big Bang Observer (BBO)~\cite{Corbin:2005ny,Crowder:2005nr}, whereas a phase transition occurring at higher energies, say $O$(10-100 TeV), would correspond to a GWS measurable by experiments such as the Einstein Telescope (ET)~\cite{Punturo:2010zz} and the Cosmic Explorer (CE)~\cite{Evans:2021gyd}. It has also been shown that an FOPT plays a crucial role in the production of dark matter~\cite{Baldes:2020kam,Baldes:2022oev}, primordial black holes~\cite{Hawking:1982ga,Khlopov:1999ys,Baker:2021nyl,Kawana:2021tde,Baker:2021sno,Hashino:2021qoq}, magnetic fields~\cite{Turner:1987bw,Baym:1995fk}, and other topological defects~\cite{Blasi:2022woz,Li:2023yzq,Blasi:2023rqi}. 

As bubbles nucleate and grow, their velocity becomes a prominent aspect of an FOPT. On one hand, the wall experiences an outward pressure because of the difference in energy densities between the unbroken and the broken phases. On the other hand, it also receives inward pressure from the particles residing in the thermal plasma. The competing influence of these two forces determines whether the wall reaches a small nonrelativistic velocity or if it continues to accelerate until an ultrarelativistic velocity is attained. Small wall velocities are well motivated in the context of electroweak baryogenesis~\cite{Vaskonen:2016yiu,Dorsch:2016nrg} and it was shown in~\cite{Fromme:2006wx} that large wall velocities would not leave enough time to generate the matter-antimatter asymmetry in front of an advancing bubble wall. However, in the recent past, it has been shown that successful generation of the baryon asymmetry can also occur with supersonic and relativistic velocities of the bubble wall~\cite{Caprini:2011uz,Cline:2020jre,Dorsch:2021ubz}. The latter possibility can be further probed as it leaves unique signatures in the GWS. 

Ultrarelativistic wall velocities can be achieved if the difference in energy ($\Delta V$) exceeds the leading order pressure ($P_{\text{LO}}$) from the thermal plasma, i.e., $\Delta V > P_{\text{LO}}$~\cite{Bodeker:2009qy,Bodeker:2017cim}. This situation has recently been realized within a BSM framework augmenting the SM fields with a gauge singlet scalar and leading to a two-step phase transition~\cite{Azatov:2022tii}. Ultrarelativistic bubbles are associated with GW spectra with high peaks~\cite{Caprini:2007xq,Huber:2008hg,Caprini:2009fx}. Recent works also indicate that both baryogenesis~\cite{Katz:2016adq,Azatov:2021irb} and dark matter production~\cite{Azatov:2021ifm,Baldes:2021vyz} can both be explained by scenarios leading to ultrarelativistic bubbles. This has enhanced the incentive for the study of new physics models that lead to such large wall velocities, for probing the physics of the electroweak phase transition at multiple frontiers. 

The prospect of FOPT has been investigated within a wide array of BSM setups. The simplest means to accommodate an FOPT is to permit significant deviation of the effective Higgs self coupling, roughly $\gtrsim 20\%$ or so~\cite{Kanemura:2004ch,Noble:2007kk,Huang:2015tdv}. This can be achieved by introducing a gauge singlet scalar field $S$, which couples to the SM Higgs field through a portal interaction. Depending on whether the singlet would acquire a vacuum expectation value or not, the phase transition occurs as a two-step or one-step phenomenon~\cite{Choi:1993cv,Espinosa:2007qk,Barger:2007im,Espinosa:2008kw,Espinosa:2011ax,Cline:2012hg,Marzola:2017jzl,Kurup:2017dzf,Schicho:2022wty}. A similar situation can occur for the well known two Higgs doublet model~\cite{Cline:1996mga,Bernon:2017jgv,Dorsch:2013wja,Dorsch:2014qja,Basler:2016obg}, triplet extended SM~\cite{Patel:2012pi,Chala:2018opy,Abdussalam:2020ssl}, radiatively generated barrier~\cite{Lofgren:2021ogg} etc. The possibility of FOPT has also been analyzed in supersymmetric models. While the stop quark induced FOPT is highly disfavored~\cite{Cohen:2012zza,Curtin:2012aa,Katz:2015uja,Liebler:2015ddv}, new parameter space opens up once the minimal version of the supersymmetric theory is extended to include new fields~\cite{Chatterjee:2022pxf,Borah:2023zsb}. 

However, a glaring lack of direct experimental evidence for new degrees of freedom, beyond the SM ones, prompts us to adopt an effective field theory (EFT) based approach. In the context of a first order EWPT, BSM effects can be neatly encapsulated within the contact interactions of the Standard Model effective field theory (SMEFT). Within this framework, several studies have explored the idea that a tree-level barrier in the potential can be generated by introducing a $H^6$ interaction, potentially resulting in FOPT~\cite{Grojean:2004xa,Huber:2007vva,Delaunay:2007wb}. Reference \cite{Cai:2017tmh} presents a detailed analysis of the predictions of GW spectra sourced by an FOPT. Additionally, it has been shown that dimension 6 structures can be introduced as new sources of CP violation to generate enhanced baryon asymmetry~\cite{Bodeker:2004ws,Zhang:1992fs}. In the SMEFT paradigm, uncertainties due to the renormalization scale dependence in the usual daisy resummation approach has been discussed in~\cite{Croon:2020cgk}. The limit of small bubble velocities in this context has been addressed using a hydrodynamic approach in \cite{Lewicki:2021pgr} [For other works pertaining to bubble wall velocities, see~\cite{Moore:1995si,Moore:1995ua,Konstandin:2010dm,BarrosoMancha:2020fay,Balaji:2020yrx,Ai:2021kak,Laurent:2022jrs}].  On the other hand, the prospect of ultrarelativistic bubbles was noted in \cite{Ellis:2018mja} with dimension 6 SMEFT operators if the new physics scale was below 580 GeV. The challenges associated with the power counting and constructing UV-completion for the SMEFT scenarios that lead to successful electroweak baryogenesis and phase transition, in general, were pointed out in Refs.~\cite{deVries:2017ncy,Postma:2020toi}. 

\begin{figure}[t]
    \centering
    \includegraphics[width=\linewidth]{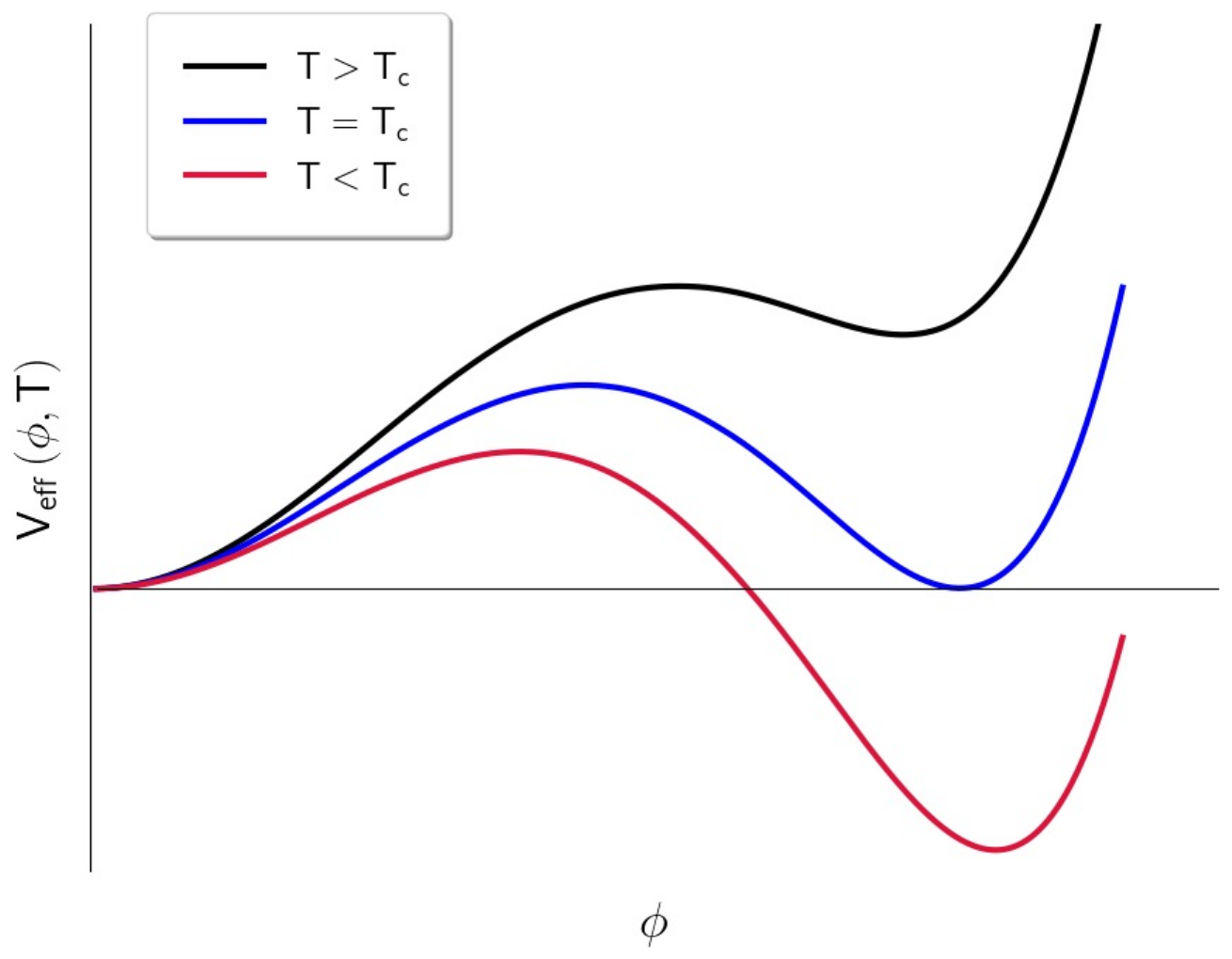}
    \caption{A schematic diagram portraying FOPT. $T_c$ is defined by the blue curve, i.e., the temperature corresponding to degenerate vacua.}
    \label{fig:fopt}
\end{figure}

The objective of our analysis is to take into account the effects of the dimension 6 SMEFT operators $\mathcal{O}_H$, $\mathcal{O}_{H\mathcal{D}}$, and $\mathcal O_{H\square}$ \cite{Grzadkowski:2010es} and scrutinize the Wilson coefficient parameter space that not only is conducive to an FOPT but also leads to ultrarelativistic bubble wall velocities. These operators directly contribute to the Higgs potential and also to the renormalization of the Higgs wave function. As a result, a consistent treatment based on renormalization group improved potential becomes necessary. Contemporary works have shed light on the parameter space for the SMEFT~\cite{Camargo-Molina:2021zgz} as well as beyond SMEFT (BSMEFT)~\cite{Anisha:2022hgv,Anisha:2023vvu} scenarios, that can lead to a strong first order phase transition based on the condition $\phi_c/T_c\gtrsim1$. Here, $T_c$ is the critical temperature, where the effective potential presents degenerate vacua, as depicted by the blue curve in Fig.~\ref{fig:fopt}, and $\phi_c$ denotes the location of the minimum of the potential at $T_c$. However, before a combination of parameter values can be declared to be suitable for a strong FOPT, it is vital to further examine the details of the bubble dynamics as the temperature gradually decreases below $T_c$. One of the aims of our work is to inspect this aspect and check whether the criterion of strong FOPT, $\phi_c/T_c\gtrsim1$, is commensurate with bubble nucleation or not, mainly focusing on ultrarelativistic bubble wall velocities. At this point, it is important to note our findings compared to Ref.~\cite{Ellis:2018gqa}. We noticed that a successful EWPT can occur only for a tiny range of $C_H$, i.e., the coefficient of the effective $(H^\dagger H)^3$ term in the Lagrangian, which for $O(1)$ coefficients translates roughly to a new physics scale between 563 and 627 GeV. These boundary values describe the limits where either the height of the barrier is too high for EWPT or the barrier would completely disappear. Moreover, ultrarelativistic bubbles are only possible around the narrow window of 563-564 GeV. However, as noted in Sec.~\ref{sec:dim8}, such low values of the cutoff scale requires further investigation with dimension 8 SMEFT operators included. Therefore, we include the most dominant contribution to the effective potential and study its impact on the nucleation temperature as well as the bubble velocity.

This article is organized as follows. We start with an in-depth discussion of the ingredients and steps involved in assembling the effective thermal potential in Sec.~\ref{sec:one-loop-effective-potential}. We describe how the zero-temperature, tree-level potential is improved through the incorporation of one-loop, SMEFT, and renormalization group (RG) effects. In Sec.~\ref{sec:bubble-profiles}, we solve for the bounce action and obtain nucleation temperatures for different values of the SMEFT coefficients. Sec.~\ref{sec:dim8} extends this analysis through the inclusion of a dimension 8 SMEFT operator. This is followed by a detailed discussion of the pressure on the bubble wall in Sec.~\ref{sec:pressure-and-velocity}. We show that bubble nucleation occurs only in specific regions of the SMEFT parameter space. In fact, the allowed ranges for the Wilson coefficients shrink considerably once the discussion is specialized toward ultrarelativistic bubble wall velocities. We evaluate the gravitational wave spectra corresponding to the parameter sets yielding bubbles with large velocities in Sec.~\ref{sec:SGWB}. Finally, we summarize our conclusions and discuss possible UV completions in Sec.~\ref{sec:conclusion}.

\section{The effective thermal potential}\label{sec:one-loop-effective-potential}
Necessary requirements for an FOPT are the existence of degenerate minima at a critical temperature $T_c$ and the appearance of a potential barrier separating the minima, at and below a critical temperature, as shown in Fig~\ref{fig:fopt}. Such a feature emerges once we extend the zero temperature potential to account for finite-temperature effects \cite{PhysRevD.7.1888,Quiros:1999jp,Curtin:2016urg,Carrington:1991hz}:
\begin{eqnarray}\label{eq:effective-potential}
    V_\text{eff}(\phi,T) &=& V_{T=0}(\phi) + V_{T\neq0}(\phi,T)\;.
\end{eqnarray}

In the subsections that follow, we systematically build the necessary ingredients of the effective potential. We extend the tree-level potential to account for one-loop effects, introduce dimension 6 SMEFT interactions, discuss the impact of renormalization group evolution (RGE) of parameters using the background field method and finally introduce the effects of finite-temperature interactions.

\subsection{Zero-temperature one-loop corrected potential}
The interactions between the Higgs scalar and other SM fields, relevant for the electroweak phase transition, can be described in terms of the following subset of the renormalizable SM Lagrangian,
\begin{eqnarray}\label{eq:scalar-lag}
\mathcal{L}(H,H^\dagger) &\supset& (\mathcal{D}_{\mu}H)^\dagger(\mathcal{D}^{\mu}H) - m_H^2\,H^\dagger H \nonumber\\
&& - \cfrac{\lambda}{2}\,(H^\dagger H)^2 - y_t\,\overline{q}_{3L}\,t_R\,\widetilde{H}\;,
\end{eqnarray}
where $\mathcal{D}_\mu$ is the covariant derivative, i.e.
\begin{eqnarray}
    \mathcal{D}_\mu\,H = \left(\partial_\mu\ + \cfrac{i}{2}\,g_2\,W^I_\mu\tau^I + \cfrac{i}{2}\,g_1\,B_\mu\right) H\;,
\end{eqnarray}
where, $g_1$, $g_2$ are the coupling constants for the $U(1)_Y$, $SU(2)_L$ gauge groups and $B_\mu$, $W^I_\mu$, ($I=1,2,3$) are the corresponding gauge bosons. $\tau^I$ are the Pauli matrices, $\widetilde{H} = i\tau_2\,H^*$, $q_{3L}$ and  $t_R$ denote the left chiral third generation quark doublet and the right chiral top quark, respectively, with $y_t$ being the corresponding Yukawa coupling.

After electroweak symmetry breaking, $W^I_\mu$, $B_\mu$ are rotated into the physical $W^\pm_\mu$, $Z_\mu$ vector bosons and the photon. The Higgs doublet can be expressed in terms of fluctuations around a background field $\langle H\rangle = \phi$ as 
\begin{eqnarray}\label{eq:matrix-form-of-phi}
 H = \begin{bmatrix}
    G^+ \\
    \cfrac{1}{\sqrt{2}}(\phi+h+iG^0)
\end{bmatrix}\;,
\end{eqnarray}
where $h$ corresponds to the dynamical Higgs field and $G^+$, $G^0$ are the Goldstone bosons. 

The effective thermal potential is a functional of the static background field which is only a function of the radial coordinate, i.e., $\phi \equiv \phi(r)$, which is defined at an arbitrary temperature and $\phi = v = 246$ GeV at $T=0$. In addition to the  terms derived from Eq.~\eqref{eq:scalar-lag}, the zero-temperature one-loop contributions, i.e. the Coleman-Weinberg corrections (in the $\overline{\text{MS}}$ renormalization scheme) \cite{PhysRevD.7.1888,Quiros:1999jp} have the following schematic form:
\begin{eqnarray}\label{eq:CW-potential}
V_\text{CW}(\phi) = \frac{1}{64\pi^2}  \,\, \sum_i \,\, n_i  \,m^4_i (\phi) \left[\, \log\,\left( \frac{m^2_i (\phi)}{\mu^2} \right) - C_i  \right],
\end{eqnarray}
where $i = W^{\pm}, Z, h, G^{\pm}, G^0, t$ refers to the individual fields and $n_i$ are the corresponding numbers of degrees of freedom. More specifically,
\begin{eqnarray}\label{eq:degrees-of-freedom}
    n_{W^\pm} &=& 6, \quad n_{Z} = 3, \quad n_{G^\pm} =2\;, \nonumber\\ n_{G^0} &=& 1, \quad n_{h} = 1, \quad n_{t} = -12\;.
\end{eqnarray}
The constant factors $C_i$ are given as 
\begin{eqnarray}
C_{W^{\pm},\, Z} = \cfrac{5}{6}  \qquad \text{and} \qquad C_{h,\, G^\pm,\, G^0,\, t,\, b} = \cfrac{3}{2}\;.  
\end{eqnarray}
After adding the Coleman-Weinberg potential, we recompute the values of the Lagrangian parameters $m_H^2, \, \lambda$ by simultaneously imposing the following renormalization conditions:
\begin{eqnarray}\label{eq:renorm-cond}
\cfrac{\partial \left(V_\text{tree} + V_\text{CW}\right)}{\partial \phi} \Bigg|_{\phi=v} &=& 0\;, \nonumber\\
\cfrac{\partial^2 \left(V_\text{tree} + V_\text{CW}\right)}{\partial \phi^2} \Bigg|_{\phi=v} &=& m^2_{h, \text{phys}} \approx (125.5 \text{ GeV})^2\;.
\end{eqnarray}
Here, $V_{\text{tree}}$ is the usual tree-level SM potential. The first of these reflects the existence of a minima of the zero-temperature potential at $\phi = v$, i.e., the vacuum expectation value at $T=0$, and the latter demands that the mass computed based on this zero-temperature potential must equal the physical mass of the Higgs scalar.

\subsection{Incorporating SMEFT (dimension 6) operators}
The SM with its specific parameter set and particle masses as affirmed by high energy experiments fails to facilitate an FOPT, despite taking into account loop and finite-temperature corrections to the scalar potential. In other words, a first order EWPT is entirely an artifact of BSM physics, and the popular frameworks that accommodate such a phenomenon consist of minimal extensions of the SM scalar sector. 

An effective field theory such as SMEFT provides an ideal backdrop that can encode the features and consequences of a variety of BSM models. In what follows, we have conducted a model-independent analysis by taking into account the effects of SMEFT operators of mass dimension 6, constituted solely of the Higgs and its derivatives. Therefore, the Lagrangian is extended to
\begin{eqnarray}
    \mathcal{L}_\text{full}(H,H^\dagger) &=& \mathcal{L}(H,H^\dagger) + \cfrac{C_H}{\Lambda^2}\,\mathcal{O}_H \nonumber\\
    && + \cfrac{C_{H\mathcal{D}}}{\Lambda^2}\,\mathcal{O}_{H\mathcal{D}} + \cfrac{C_{H\square}}{\Lambda^2}\,\mathcal{O}_{H\square}\;,
\end{eqnarray}
where $\mathcal{O}_H$, $\mathcal{O}_{H\mathcal{D}}$, and $\mathcal{O}_{H\square}$ denote the  operators,
\begin{align}
    \mathcal{O}_H = ( & H^\dagger H)^3\;,\; \;  \; \;\;\mathcal{O}_{H\mathcal{D}} = (H^{\dagger}i\overleftrightarrow{\mathcal{D}_\mu}H)^2\;, \nonumber\\
   & \mathcal{O}_{H\square} = (H^\dagger H)\square(H^\dagger H)\;,
   \label{eq:operators}
\end{align}
and $C_H$, $C_{H\mathcal{D}}$, and $C_{H\square}$ are the corresponding Wilson coefficients (WCs). $\Lambda$ refers to the unknown high energy scale but for a major portion of our discussion, we will absorb it within the definition of the WCs, i.e., $C_i/\Lambda^2\rightarrow C_i$. $C_H$ participates directly in the tree level Lagrangian, and after symmetry breaking we get,
\begin{eqnarray}\label{eq:tree-level-potential}
    V_\text{tree}(\phi) = \frac{1}{2} m^2_H\,\phi^2 + \frac{1}{8}\lambda\,\phi^4 - \frac{1}{8}C_H\,\phi^6\;,
    \label{eq:tree_level_pot}
\end{eqnarray}
where $m_H$ is the Higgs mass parameter in the tree-level Lagrangian.

$C_{H\mathcal{D}}$ and $C_{H\square}$, on the other hand, offer modifications to the kinetic term of the physical scalar ($h$). This necessitates a field redefinition of the form $h \rightarrow Z_h^{-1}\,h$, so as to retrieve the canonical form of the kinetic term, with
\begin{eqnarray}\label{eq:field-redefinition}
    Z_h = (1-C_{H\square}\,v^2+\cfrac{1}{4}\,C_{H\mathcal{D}}\,v^2)\;.
\end{eqnarray}
Once again, $v$ refers to the vacuum expectation value at $T=0$. While the static background field $\phi(r)$ does not undergo the same field redefinition, the effect of Eq.~\eqref{eq:field-redefinition} is captured within the expressions for the field-dependent masses of the various degrees of freedom, collected in Eq.~\eqref{eq:field-dep-masses}.

\begin{eqnarray}\label{eq:field-dep-masses}
    m^2_{W}(\phi) &=& \cfrac{1}{4}\,g_2^2\phi^2\;, \nonumber\\ 
    m^2_{Z}(\phi) &=& \cfrac{1}{4}\,\left(\cfrac{3}{5}g_1^2+g_2^2\right)\phi^2 \,(1+\cfrac{1}{2}\,C_{H\mathcal{D}}\,\phi^2)\;, \nonumber\\
    m^2_{G^\pm}(\phi) &=& m_H^2+\cfrac{\lambda}{2}\phi^2 - \cfrac{3}{4}\,C_H\,\phi^4\;, \nonumber\\
    m^2_{G^0}(\phi) &=& m_H^2+\cfrac{\lambda}{2}\phi^2 -\cfrac{m_H^2}{2}\,C_{H\mathcal{D}}\,\phi^2 \nonumber\\
    &&-\cfrac{3}{4}\,C_H\,\phi^4-\cfrac{\lambda}{4}\,C_{H\mathcal{D}}\,\phi^4\;, \nonumber\\
    m^2_h(\phi) &=& m_H^2+\cfrac{3\lambda}{2}\phi^2 -\cfrac{m_H^2}{2}\,(C_{H\mathcal{D}}-4\,C_{H\square})\,\phi^2 \nonumber\\
    &&-\cfrac{3}{4}\,(5\,C_H+\lambda\,(C_{H\mathcal{D}}-4\,C_{H\square}))\,\phi^4\;,\nonumber\\
    m^2_t(\phi) &=& \cfrac{1}{2}\,y_t^2\,\phi^2.
\end{eqnarray}

\subsection{Renormalization group improved potential}
The effective potential in Eq.~\eqref{eq:CW-potential} involves the renormalization scale $\mu$ which is not physical. Our construction should be independent of $\mu$ and this can be achieved by constructing a RG improved effective potential. This would ensure that a change in $\mu$ is accompanied by the change in renormalized parameters. In order to do this, we use the background field method~\cite{Kastening:1991gv,Ford:1992mv,Manohar:2020nzp} to evaluate the RG evolution equations. The principal idea is to write the Higgs field in terms of a fluctuation and a slowly varying background field $\phi$. This decomposition is analogous to the sharp momentum cutoff scheme. The effective potential, composed of both tree-level and one-loop CW terms, is now written in terms of the background field as shown in Eqs.~\eqref{eq:tree_level_pot} and \eqref{eq:CW-potential}, while the fluctuation is integrated over in the functional integral. The all order zero-temperature potential obeys the RG equation
\begin{eqnarray}\label{eq:CS-equation}
    \left(\cfrac{\partial}{\partial t} + \beta_i\,\cfrac{\partial}{\partial \lambda_i} - \gamma_\phi\,\phi\,\cfrac{\partial}{\partial \phi}\right)V(\phi_{\text{cl}}, \lambda_i, \mu) = 0\;.
\end{eqnarray}
Here, $\lambda_i$ is the set of Lagrangian parameters such as $m_H^2$, $\lambda$, $C_H$, and so on. The solution of Eq~\eqref{eq:CS-equation} is also well-known where the dependence of the sliding energy scale $\mu$ is described completely by the running parameters, i.e.,
\begin{equation}
    V(\phi_{\text{cl}}, \lambda_i, \mu) = V(\overline{\phi}_{\text{cl}}(\mu^\prime), \overline{\lambda_i}(\mu^\prime), \mu^\prime)\;,
\end{equation}
where the barred quantities satisfy their corresponding $\beta$ functions given by
\begin{eqnarray}
    \cfrac{d\overline{\lambda_i}}{dt} = \beta_i \{\overline{\lambda_j}\}, \qquad t = \cfrac{1}{16\pi^2}\,\ln{\cfrac{\mu}{\mu_0}}\;,
\end{eqnarray}
with $\mu$ being the scale of renormalization and $\mu_0$ being the input or reference scale. We can further simplify Eq.~\eqref{eq:CS-equation} as the tree-level potential does not depend on the renormalization scale:
\begin{equation}\label{eq:simplified-CS-equation}
    \left( \beta_i\,\cfrac{\partial}{\partial \lambda_i} - \gamma_\phi\,\phi\,\cfrac{\partial}{\partial \phi}\right)V_{\text{tree}}(\phi_{\text{cl}}, ..) = - \,\cfrac{\partial}{\partial t} \,V_{\text{CW}}(\phi_{\text{cl}}, ..)\;. 
\end{equation}
After incorporating the explicit expressions of $V_{\text{tree}}$ and $V_{\text{CW}}$ into Eq.~\eqref{eq:simplified-CS-equation}, we can read off the $\beta$ functions for each of the Lagrangian parameters by equating the  coefficients of $\phi^2$, $\phi^4$, and $\phi^6$ on both sides of Eq.~\eqref{eq:simplified-CS-equation}. This results in
\begin{eqnarray}\label{eq:beta-func-1}
\beta_{m_H^2}&\simeq&\cfrac{1}{16\pi^2} \bigg[2\,m_H^2\,\gamma_{\phi} + 6\,\lambda\, m_H^2\bigg]\;,\nonumber\\
\beta_{\lambda}&\simeq&\cfrac{1}{16\pi^2}\bigg[4\,\lambda\,\gamma_{\phi} + 12\,\lambda^2 - 48\,C_Hm_H^2-12\,y_t^4\bigg]\;,\nonumber\\
\beta_{C_H} &\simeq&\frac{1}{16\pi^2}\bigg[54\,C_H \lambda  + 6\, C_H \gamma_{\phi}\bigg]\;,
\end{eqnarray}
where $\gamma_\phi=3y_t^2$ is the anomalous dimension of the field $\phi$. Note that, at one-loop, the running of $C_{H\square}$ and $C_{H\mathcal{D}}$ is self-proportional and therefore we ignore their contributions~\footnote{Our results are in agreement with the RGEs obtained in \cite{Jenkins:2013zja,Alonso:2013hga,Manohar:2020nzp}. The difference in some coefficients can be attributed to the way couplings are defined in the potential.}. However, they will contribute in the physical masses of the fields as shown in Eq.~\eqref{eq:field-dep-masses}. Similarly, $y_t$ and $g_3$ follow their usual RGEs from the wave function renormalization pieces
\begin{eqnarray}\label{eq:beta-func-2}
\beta_{y_t}&\simeq&\cfrac{1}{16\pi^2}\bigg[\cfrac{9}{2}\,y_t^3-8\,g_3^2\, y_t \bigg]\;,\nonumber\\
\beta_{g_3}&\simeq&\cfrac{1}{16\pi^2}\bigg[-7 g_3^3\bigg]\;.
\end{eqnarray}

With these RG flows of the parameters, the renormalization group improved effective potential can be obtained by choosing $\mu\to v$ and replacing all the parameters with renormalized parameters in Eq.~\eqref{eq:CW-potential}. This resums all the large logarithms present in the CW potential. Notice that, such a method can be extended to multiscale problems as well, where arbitrary choices of the renormalization scale might not minimize all the large logarithms. The RG flows of the parameters also help us to match our WCs at the scale $\Lambda$, where all the operators in Eq.~\eqref{eq:operators} are generated. We will quantify the impact of the RG improved potential on nucleation temperature and bubble dynamics in Secs.~\ref{sec:bubble-profiles} and \ref{sec:pressure-and-velocity}, respectively.

\subsection{Finite temperature corrections}
Temperature-dependent corrections to the effective potential can be summarized as follows \cite{Quiros:1999jp,Carrington:1991hz}
\begin{eqnarray}\label{eq:finite-T-potential}
V_\text{finite-T}(\phi,T) = \cfrac{T^4}{2\,\pi^2} \, \left[\sum\limits_{i}\, n_i \,J_{B/F} \left(\cfrac{m^2_i(\phi)}{T^2}\right)  \right]\;,
\end{eqnarray}
where $i$ once again, denotes the fields, $n_i$ represents the numbers of degrees of freedom as mentioned in Eq.~\eqref{eq:degrees-of-freedom} and the corresponding masses $m_i^2(\phi)$ are given in Eq.~\eqref{eq:field-dep-masses}. The thermal functions $J_{B,F}$ correspond to bosonic and fermionic degrees of freedom and these can be expressed as the following integrals,
\begin{equation}
J_{B,F}\left(\cfrac{m^2}{T^2}\right) = \int\limits_{0}^\infty \, dx\, x^2\, \log \left[ 1 \mp e^{ -\sqrt{x^2 + m^2/T^2} } \right]\;.
\end{equation} 
It is well-known that, at the critical temperature, the one-loop approximation mentioned before breaks down. Daisy resummation ensures that all the IR divergent pieces are resummed in the following manner
\cite{Curtin:2016urg,Carrington:1991hz,Quiros:1992ez}:
\begin{eqnarray}\label{eq:daisy-term}
V_\text{daisy}(\phi,T) = -\cfrac{T}{12\pi}\,\sum_{i}\left[ m^3_i(\phi,T) - m^3_i(\phi,0)\right]\;,
\end{eqnarray}
where $i$ now refers to $h$,  $G^\pm, G^0$, and the longitudinal modes of the vector bosons $W^\pm_l$, $Z_l$, $\gamma_l$. Note that, adding Eq.~\eqref{eq:daisy-term} is equivalent to the substitution $m_i^2\to m_i^2+\Pi_i$ in the effective potential, where $\Pi_i$ is the leading self energy corrections corresponding to the one-loop thermal mass. However, we use Eq.~\eqref{eq:daisy-term} in the effective potential and take into account the full thermally corrected potential instead of the high temperature approximation. The temperature-dependent masses can be obtained as
\begin{eqnarray}
    m^2_{W,l}(\phi,T) &=& m^2_W(\phi)+\cfrac{11}{6}\,g_2^2T^2\;, \nonumber\\
    m^2_{Z,l}(\phi,T) &=& \cfrac{1}{2}\left[m^2_{Z}(\phi)+\cfrac{11g_2^2T^2}{6}+\cfrac{11g_1^2T^2}{10}+\Delta(\phi,T)\right]\;, \nonumber\\
    m^2_{\gamma,l}(\phi,T) &=& \cfrac{1}{2}\left[m^2_{Z}(\phi)+\cfrac{11g_2^2T^2}{6}+\cfrac{11g_1^2T^2}{10}-\Delta(\phi,T)\right]\;, \nonumber\\
    m^2_{i}(\phi, T) &=& m^2_{i}(\phi) + \left(\cfrac{\lambda}{4} + \cfrac{3g_1^2}{80}+\cfrac{3g_2^2}{16}+\cfrac{y_t^2}{4}\right)\,T^2\;,\nonumber\\
\end{eqnarray}
with $i$ = $h, G^0, G^\pm$ and
\begin{eqnarray}
    \Delta(\phi,T) &=& \sqrt{m^4_Z(\phi)+\cfrac{11 T^2}{900} \left(3 g_1^2-5 g_2^2\right)^2 \left(3  \phi^2+11  T^2\right)}\;.\nonumber
\end{eqnarray}

The $T=0$ as well as $T \neq 0$ pieces of the total thermal potential can be summarized as
\begin{eqnarray}
    V_{T=0}(\phi) &=& V_\text{tree}(\phi) + V_\text{CW}(\phi)\;, \nonumber\\
    V_{T\neq0}(\phi,T) &=& V_\text{finite-T}(\phi,T) + V_\text{daisy}(\phi,T)\;.
\end{eqnarray}

\newpage
\section{Bubble profiles and the nucleation temperature}\label{sec:bubble-profiles}

The effective potential exhibits degenerate minima at the critical temperature, and a non-zero probability for tunneling between the false and true vacua exists for temperatures below $T_c$. But for the phase transition to succeed the system must cool below the nucleation temperature $T_{\text{nucl}}$ where the probability of nucleation of a single bubble per 
Hubble volume per Hubble time is $O(1)$.
\begin{eqnarray}\label{eq:nucleation-def-1}
P\,(T_{\text{nucl}}) = \int \limits_{\infty}^{T_{\text{nucl}}}\,\,\cfrac{dT}{T}\left( \cfrac{2\,\xi M_{\text{Pl}}}{T}\right)^4\,e^{-S_3(T)/T} \simeq 1\;.
\end{eqnarray} 
Here, $M_{\text{Pl}}$ is the Planck mass scale and $\xi = 4\pi \sqrt{\pi\,g_*(T)/45}$, with $g_*(T_{\text{nucl}}) \sim 100$ denoting the number of relativistic degrees of freedom in the thermal plasma. The criteria in Eq.~\eqref{eq:nucleation-def-1} can be rewritten in a simplified form as
\begin{eqnarray}\label{eq:nucleation-def-2}
    \frac{S_3(\phi(r), T) - S_3(0, T)}{T}\bigg|_{T=T_\text{nucl}} \simeq 140\;,
\end{eqnarray}
where $S_3$ is the classical bounce action in 3 dimensions,
\begin{eqnarray}\label{eq:action}
    S_3(\phi, T) = 4\pi\,\int_0^\infty\, dr\,r^2 \left[\left(\cfrac{d\phi}{dr}\right)^2 + V_\text{eff}(\phi,T)\right]\;.
\end{eqnarray}

\begin{figure}[ht]
    \centering
    \includegraphics[width=\linewidth]{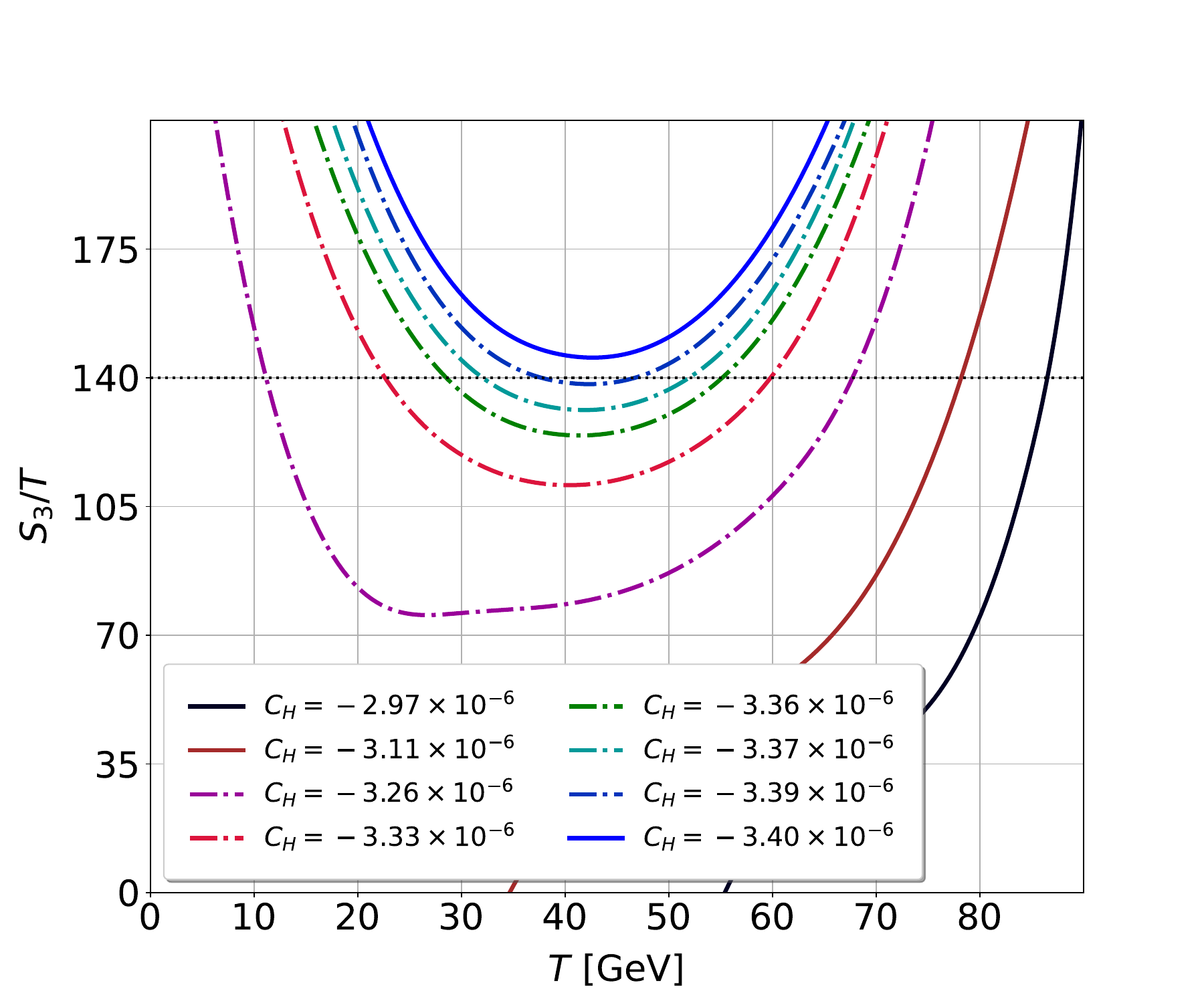}
    \caption{The effect of varying $C_H$ on the ratio $S_3/T$. Each line in the figure corresponds to a specific $C_H$ value (expressed in units of GeV${}^{-2}$). The dashed line denotes the criteria mentioned in Eq.~\eqref{eq:nucleation-def-2}, i.e. $S_3/T = 140$. For each case, we have set $C_{H\square}$ =$1.56\times 10^{-7}$ GeV${}^{-2}$ and $C_{H\mathcal{D}}$ = $-1.15\times 10^{-8}$ GeV${}^{-2}$.}
    \label{fig:S3/T_vs_T}
\end{figure}

Tunneling from the false vacuum to the true vacuum below the critical temperature is an instantaneous process and the trajectory for the same is dictated by the static bubble profiles which are defined for fixed time slices and at a particular temperature. Therefore, these are simply a function of the radial coordinate. The profiles  $\phi(r)$ can be obtained as a solution of the equation of motion:
\begin{eqnarray}
    \partial^2 \phi + \cfrac{\partial V_\text{eff}(\phi,T)}{\partial \phi} = 0\;.
\end{eqnarray}
This can be recast as the following initial value problem,
\begin{eqnarray}\label{eq:DE}
-\cfrac{1}{r^2}\,\cfrac{\partial}{\partial\,r}\,\left(r^2\,\cfrac{\partial\, \phi}{\partial \,r}\right) + \cfrac{\partial\,V_{\text{eff}}(\phi, T)}{\partial\, \phi} = 0\;,
\end{eqnarray}
subject to the boundary conditions
\begin{eqnarray}\label{eq:DE-bcs}
\phi^\prime(r) \xrightarrow{r \rightarrow 0} \,0, \hspace{0.3cm} \phi^\prime(r) \xrightarrow{r \rightarrow \infty} \,0, \hspace{0.3cm} \phi(r) \xrightarrow{r \rightarrow \infty} \,0\;.
\end{eqnarray}

We find solutions for the single-field system described in Eq.~\eqref{eq:DE} using a numerical approach based on the shooting method for solving initial value problems. The basic idea behind this approach is to pinpoint an initial field configuration $\phi(0)$, such that if a particle starts with a vanishing gradient near the maxima of an inverted potential, i.e., $\phi^\prime(0) = 0$, then it would arrive at the configuration corresponding to the false vacuum at large $r$ with a vanishing gradient, i.e., $\phi^\prime(r) = 0$ when $\phi(r) = 0$ for $r\rightarrow\infty$. To obtain the solutions of Eq.~\eqref{eq:DE} for constant values of $T$, we relied on a simple mathematica based implementation of the shooting method~\footnote{A variety of sophisticated computations tools such as \texttt{CosmoTransitions} \cite{Wainwright:2011kj}, \texttt{FindBounce} \cite{Guada:2020xnz}, \texttt{Elvet} \cite{Araz:2021hpx} can also be utilized to obtain the bubble profiles. On the other hand \texttt{DRalgo}~\cite{Ekstedt:2022bff} computes finite temperature effective potential within a effective theory.}. Having obtained the solutions $\phi \equiv \phi(r)$ at different temperatures, we then used Eq.~\eqref{eq:nucleation-def-2} to ascertain the nucleation temperature.

In the process of determining the static solution $\phi(r)$, we rescaled all dimensionful quantities in Eq.~\eqref{eq:DE} with respect to the mass of the $W$-boson, $m_W = 80$ GeV, i.e. we multiply and divide Eq.~\eqref{eq:DE} by powers of $m_W$ and make the following identifications,
\begin{eqnarray}\label{eq:rescaling}
&& \cfrac{1}{m_W}\,\, \phi\rightarrow \phi\;, \qquad \cfrac{1}{m_W}\,\, T \rightarrow T\;,\nonumber\\ 
&& m_W\, r \rightarrow r\;,\qquad \cfrac{1}{m_W}\,\, \partial_r \rightarrow \,\partial_r\;.
\end{eqnarray}
This also implies a rescaling of dimensionful parameters in the effective potential, i.e.,
\begin{eqnarray}\label{eq:rescaling1}
&& \cfrac{m_H^2}{m_W^2}\,\rightarrow \, m_H^2\;,\qquad m^2_W\, C_i \rightarrow C_i\;,
\end{eqnarray}
with $C_i \in \{C_H, C_{H\mathcal{D}}, C_{H\square}\}$.

\subsection{Parameter selection and results}

In our numerical analysis, we treat $C_H$, $C_{H\square}$, $C_{H\mathcal{D}}$ as free and independent parameters. Our choice of parameter ranges is informed by the results of recent global fits, conducted on SMEFT WCs, based on~\cite{Ellis:2018gqa,Dawson:2020oco}. These global fits were obtained by incorporating Higgs data from $8$ and $13$ TeV runs at the LHC as well as diboson processes ($WW$ and $WZ$) and electroweak precision observables including the $W$-boson mass and decay width from LEP-2 data. The resulting limits can be summarized (in units of GeV$^{-2}$) as
\begin{eqnarray}\label{eq:constraints}
    C_H &\in& [-2\times 10^{-5},\, 0.5\times10^{-5}]\;, \nonumber\\
    C_{H\square} &\in& [-0.4\times 10^{-6},\, 0.5\times 10^{-6}]\;, \nonumber\\
    C_{H\mathcal{D}} &\in& [-0.25\times 10^{-7},\, -0.02\times 10^{-7}]\;.
\end{eqnarray}

Selection of benchmark points for our analysis was done by fixing the WC values as well as $m^2_H$ and $\lambda$, in accordance with Eq.~\eqref{eq:renorm-cond} at the energy scale of 1 TeV and using RGE to evaluate the corresponding values at the electroweak scale, where the relevant SM parameters assume the following values. Here, we have employed grand unified theory normalization for $g_1$, i.e. $g_1\rightarrow\sqrt{3/5}\,g_1$.
\begin{eqnarray}
&& g_1 = 0.46\;, \quad g_2 = 0.65\;, \quad y_t = 0.91\;,
\end{eqnarray}

For a particular choice of the SMEFT WCs, the critical temperature $T_c$ can be determined based on the features of the effective thermal potential, i.e., it is the temperature where potential exhibits degenerate minima, as shown in Fig.~\ref{fig:fopt}. 
For temperatures below $T_c$, the bubble profiles $\phi(r)$ can be obtained by solving Eq.~\eqref{eq:DE}, and the ratio $S_3[\phi(r),T]/T$ can then be computed for each distinct choice of parameters. We have shown the variation in $S_3[\phi(r),T]/T$ with respect to $T$, for different choices of $C_H$ and fixed values of $C_{H\square}$, $C_{H\mathcal{D}}$ in Fig.~\ref{fig:S3/T_vs_T}.

\begin{figure}[!htb]
    \centering
    \includegraphics[width=\linewidth]{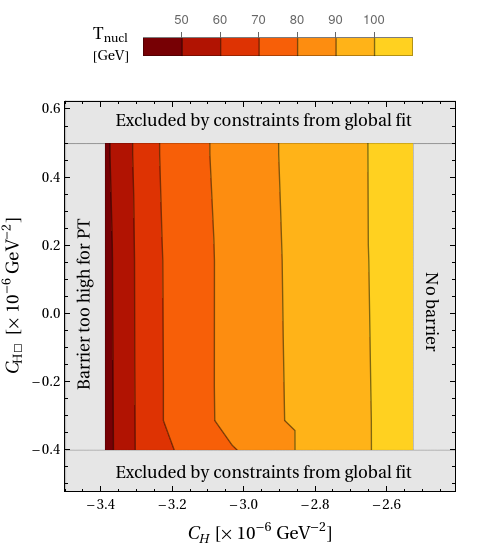}
    \caption{The variation of the nucleation temperature $T_{\text{nucl}}$ with respect to the simultaneous changes in $C_H$ and $C_{H\square}$. The parameter space excluded by various constraints has been indicated by the gray-shaded region. For $C_H > -2.53 \times 10^{-6} $ GeV$^{-2}$, the barrier separating the two minima disappears, whereas for $C_H < -3.39 \times 10^{-6}$ GeV$^{-2}$, the barrier is too high for the bubbles to nucleate. Here, we have set $C_{H\mathcal{D}} = -1.15\times10^{-8}$ GeV$^{-2}$ for each combination of $\left(C_H, C_{H\square}\right)$ values.}
    \label{fig:chbox_vs_ch}
\end{figure}

The process of selecting benchmark points also brought to light the following noteworthy observations:
\begin{enumerate}
    \item Among the three dimension 6 parameters, $C_H$ has the most notable impact on the features of the effective thermal potential and consequently on the nucleation temperatures $T_\text{nucl}$, whereas the impact of $C_{H\mathcal{D}}$ is negligibly small. We highlight the relative impact of $C_H$ vs $C_{H\square}$ on $T_\text{nucl}$ in Fig.~\ref{fig:chbox_vs_ch}. Due to the relatively smaller effects of the variation in  $C_{H\square}$ and $C_{H\mathcal{D}}$ on the nucleation temperature, we set these coefficients to fixed values for the rest of our analysis and focus mainly on the impact of $C_H$. 

    \item Not the entire range of allowed WC values, determined using global fits, is conducive to a first order EWPT. We observed that for $C_H > -2.53 \times 10^{-6}$ GeV$^{-2}$ (at the electroweak scale), a potential barrier never forms, there are no degenerate minima and the phase transition is a smooth crossover. On the other hand, for $C_H < -3.39 \times 10^{-6}$ GeV$^{-2}$, the height of the barrier remains significantly large even as $T\rightarrow 0$ and nucleation never happens. For order one coefficients at the TeV scale, these upper and lower bounds on $C_H$ correspond to an allowed range of 563.57-627.88 GeV for $\Lambda$. 
    
\end{enumerate}
Therefore, it is safe to say that EWPT offers much improvement of the constraints on the relevant WCs, as compared to the ones obtained solely based on collider data. We have catalogued the benchmark points selected for further analysis in Table~\ref{table:benchmark-points}. For each case, we have also listed the corresponding $T_c$ and $T_\text{nucl}$ values. 

\begin{figure}[!htb]
    \centering
    \includegraphics[width=\linewidth]{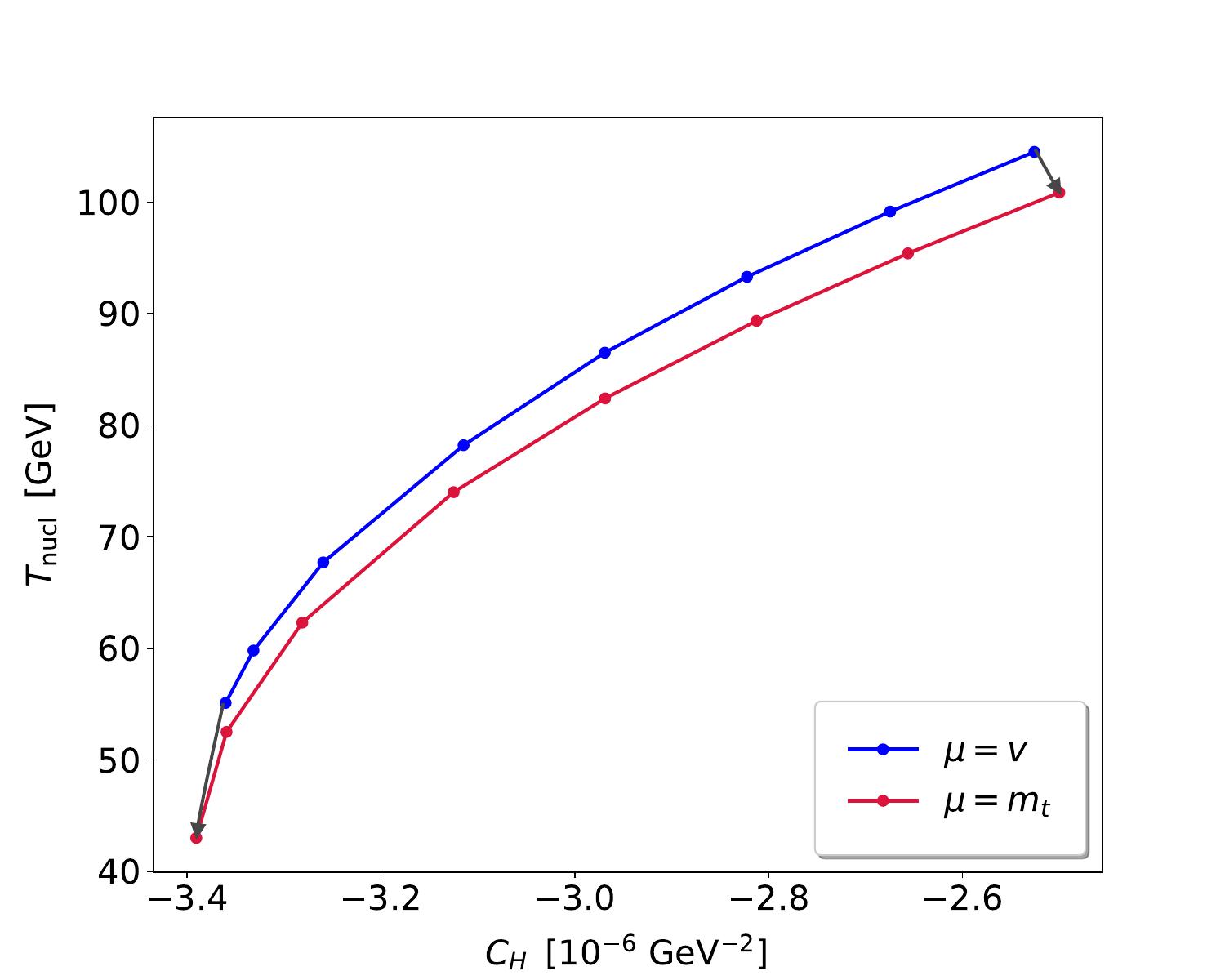}
    \caption{The impact of RGE of running parameters on $C_H$ and the corresponding $T_\text{nucl}$ values. For the purposes of our analysis, we set the renormalization scale $\mu = v \simeq 246$ GeV, i.e., the vacuum expectation value of the SM Higgs, within the Coleman-Weinberg potential, and the parameters described in Eqs.~\eqref{eq:beta-func-1} and \eqref{eq:beta-func-2} get replaced by their running counterparts. The $(C_H, T_\text{nucl})$ values with this setup are described using the blue line. The red
    line shows how these values change when the parameters are run down to the top quark mass scale $\mu = m_t \simeq 172$ GeV. We note that the changes in $T_\text{nucl}$ can be as small as $4\%$ (on the far right) and as large as $20\%$ (on the far left).}
    \label{fig:rge-Tnucl}
\end{figure}

Through a simple exercise of RG evolving the running parameters from the electroweak scale $\mu = v \simeq 246$ GeV down to the top quark mass scale $\mu = m_t \simeq 172$ GeV, we noticed differences in $T_\text{nucl}$ as small as $\sim 4\%$ and as large as $\sim 20\%$ for the benchmark points. This has been elucidated in Fig.~\ref{fig:rge-Tnucl}. While we have restricted our discussion to the two aforementioned distinct choices for $\mu$, Ref.~\cite{Croon:2020cgk} discusses in great detail what constitutes an optimal choice for $\mu$, especially when it is expressed as a linear function of the temperature characterizing the phase transition. Selecting the optimal linear relation between $\mu$ and $T$, i.e., $\mu = 2.2\,T$, see Ref.~\cite{Croon:2020cgk}, we found the values of $T_\text{nucl}$ to lie within the region between the two curves in Fig.~\ref{fig:rge-Tnucl}.

\section{Impact of dimension 8 operators}
\label{sec:dim8}
\begin{figure}[ht]
    \centering
    \includegraphics[width=\linewidth]{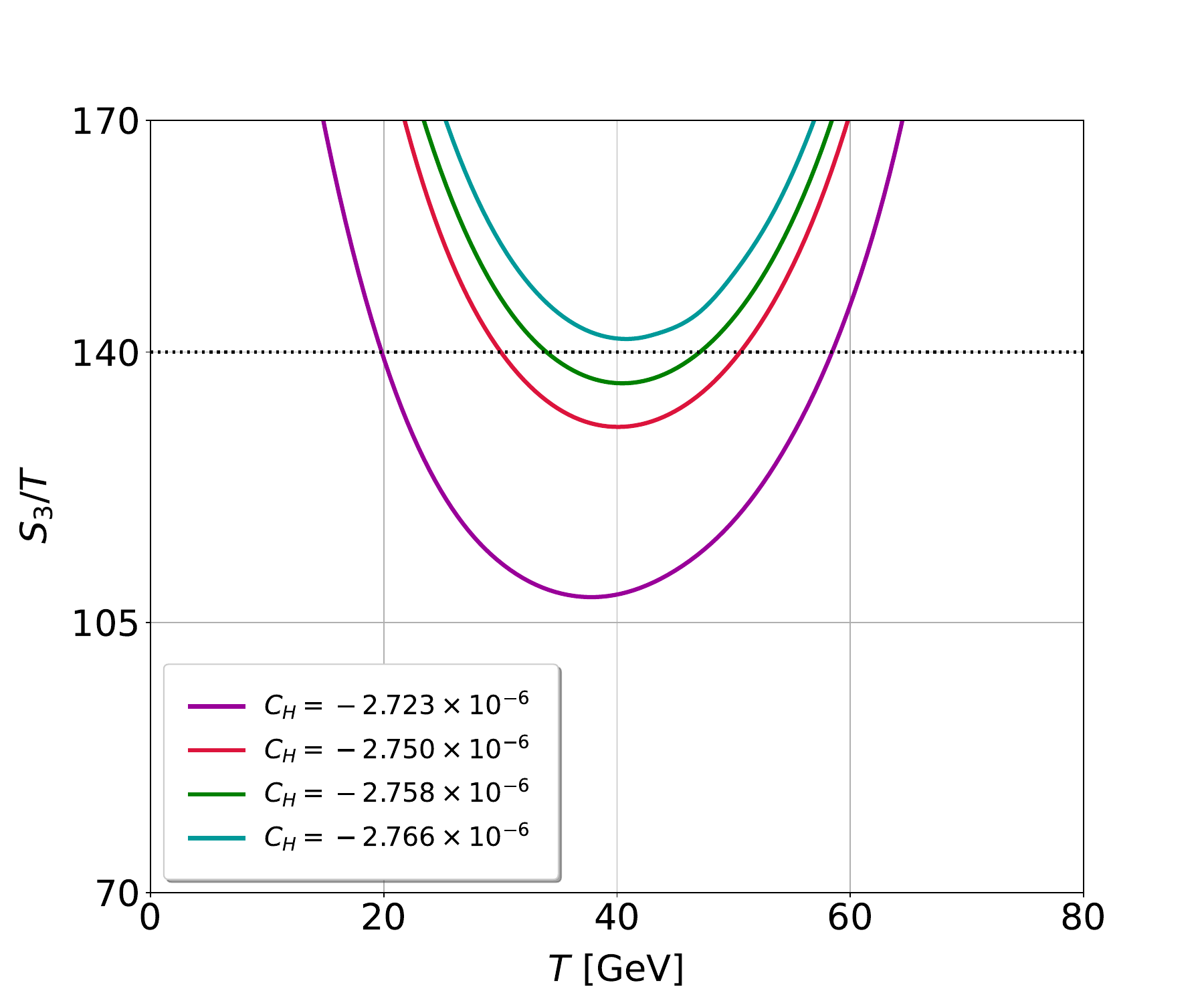}
    \caption{The effect of varying $C_H$ on the ratio $S_3/T$ in the presence of a dimension 8 term in the effective potential. Each line in the figure corresponds to a specific $C_H$ value (expressed in units of GeV${}^{-2}$). The dashed line denotes the criteria mentioned in Eq.~\eqref{eq:nucleation-def-2}, i.e. $S_3/T = 140$. For each case, we have set $C_{H\square}$ =$1.56\times 10^{-7}$ GeV${}^{-2}$ and $C_{H\mathcal{D}}$ = $-1.15\times 10^{-8}$ GeV${}^{-2}$.}
    \label{fig:S3/T_vs_T_d8}
\end{figure}

As mentioned in the previous section, the low ranges (563 GeV $\lesssim \Lambda \lesssim$ 630 GeV) for the new physics scale required for FOPT necessitate a careful analysis of the effective potential in the presence of dimension 8 operators. We first look at the power counting of the effective potential in the presence of a dimension 8 term, $C_{H^8}/16$. The barrier height is chiefly dictated by the first derivative of the potential. Taking into account the most dominant effect, i.e., top loop contribution on the Coleman-Weinberg potential, we find 
\begin{align}
    & m_H^2 + \frac{\lambda}{2}\phi^2 + \frac{\phi^2}{2\pi^2}  \log\left(\frac{m_t^2}{\mu^2}\right) - \frac{3C_H}{4}\phi^4 \nonumber \\
    &+ \frac{C_{H^8}}{2}\phi^6=0\;.
\end{align}
Arguing that the contribution from the CW should dominate over the dimension 8 term, we get $\Lambda\gtrsim 1.7 v$. This limit ensures a viable power counting of the effective potential. However, it is quite clear the dimension 8 operators would provide a sizable contribution to our analysis. In this light, we first incorporate a dimension 8 scalar self-interaction term so that the tree-level potential of Eq.~\eqref{eq:tree-level-potential} gets modified to
\begin{eqnarray}
    V^\text{d8}_\text{tree}(\phi) = V_\text{tree}(\phi) + \frac{1}{16}C_H^2\,\phi^8,
\end{eqnarray}
where once again the powers of the UV scale $\Lambda$ are absorbed within the definition of the Wilson coefficient. Also, to maintain a correlation between the dimension 6 and dimension 8 Wilson coefficients, we have set $ C_{H^8}/\Lambda^4 \equiv \left(C_H/\Lambda^2\right)^2$. 
In addition to modifying the tree-level potential, the dimension 8 operator also modifies the field-dependent masses of the Higgs as well as the Goldstones. Following are the updated expressions, 
\begin{eqnarray}\label{eq:field-dep-masses-dim8}
    m^2_{G^\pm}(\phi) &=& m_H^2+\cfrac{\lambda}{2}\phi^2 - \big(\,\cfrac{3}{4}\,C_H - \cfrac{1}{2}\,C_H^2\big)\,\phi^4\;, \nonumber\\
    m^2_{G^0}(\phi) &=& m_H^2+\big(\,\cfrac{\lambda}{2} -\cfrac{m_H^2}{2}\,C_{H\mathcal{D}}\big)\,\phi^2 \nonumber\\
    &&-\big(\cfrac{3}{4}\,C_H+\cfrac{\lambda}{4}\,C_{H\mathcal{D}}-\cfrac{3}{16}\,m_H^2\,C_{H\mathcal{D}}^2\big)\,\phi^4\nonumber\\
    &&+\big(\cfrac{3}{8}\,C_H\,C_{H\mathcal{D}}+\cfrac{1}{2}\,C_H^2+\cfrac{3}{32}\,\lambda\,C_{H\mathcal{D}}^2\big)\,\phi^6\;, \nonumber\\
    m^2_h(\phi) &=& m_H^2+\cfrac{3\lambda}{2}\phi^2 -\cfrac{m_H^2}{2}\,(C_{H\mathcal{D}}-4\,C_{H\square})\,\phi^2 \nonumber\\
    &&-(\cfrac{15}{4}\,C_H+\cfrac{3}{4}\,\lambda(C_{H\mathcal{D}}-4\,C_{H\square})-3m_H^2C_{H\square}^2\nonumber\\
    &&+\cfrac{3}{2}m_H^2\,C_{H\square}\,C_{H\mathcal{D}}-\cfrac{3}{16}m_H^2\,C_{H\mathcal{D}}^2)\,\phi^4 \nonumber\\
    &&+\big(\cfrac{7}{2}\,C_H^2-\cfrac{15}{2}\,C_H\,C_{H\square}+\cfrac{15}{8}\,C_H\,C_{H\mathcal{D}}\nonumber\\
    &&+\cfrac{9}{2}\,\lambda\,C_{H\square}^2-\cfrac{9}{4}\,\lambda\,C_{H\mathcal{D}}\,C_{H\square}+\cfrac{9}{32}\,\lambda\,C_{H\mathcal{D}}^2\big)\,\phi^6.\nonumber\\
\end{eqnarray}
O($1/\Lambda^4$) corrections to the SM mass spectrum due to SMEFT operators have also been computed in Ref.~\cite{Hays:2018zze}. With the changes in the mass spectrum as presented in Eq.~\eqref{eq:field-dep-masses-dim8}, we once again attempt to solve Eq.~\eqref{eq:DE} to obtain bubble profiles at fixed temperatures and utilize the criterion sketched in Eq.~\eqref{eq:nucleation-def-2} to ascertain the nucleation temperatures for a given set of parameters. We have catalogued benchmark points of interest, demarcated by the specific choice of $C_H$ in Table~\ref{table:benchmark-points-with-dim8}. Similar to the contents of Table~\ref{table:benchmark-points}, we have listed the $C_H$ values at the electroweak scale as well as at $\Lambda = 1$ TeV. We have also highlighted the $T_c$, $\phi_c/T_c$, and $T_\text{nucl}$ for each case. The first and last rows of Table~\ref{table:benchmark-points-with-dim8} correspond to the two limiting cases where (i) for $C_H > -1.911 \times 10^{-6}$ GeV${}^{-2}$, barrier formation does not occur, hence ruling out an FOPT and (ii) for $C_H < -2.766 \times 10^{-6}$ GeV${}^{-2}$, a sizeable barrier height even as $T \rightarrow 0$ implies that the nucleation criteria is never satisfied. A noteworthy observation when comparing the contents of Tables~\ref{table:benchmark-points} and \ref{table:benchmark-points-with-dim8} is the overall shift in the $C_H$ values and consequently in the values of the UV scale $\Lambda$, which now falls within the range of 619.16-708.58 GeV. We have shown the variation in $S_3[\phi(r),T]/T$ with respect to $T$ for the last four entries of Table~\ref{table:benchmark-points-with-dim8} in Fig.~\ref{fig:S3/T_vs_T_d8}.

\section{Ultrarelativistic bubbles}\label{sec:pressure-and-velocity}

\begin{table*}[t]
    \centering
    \setlength{\tabcolsep}{12pt}
    \renewcommand{\arraystretch}{1.9}
    \begin{tabular}{|c|c|c|c|c|c|c|c|c|c|}
         \hline
         \hline
         $C_H\times 10^{6} $&$C_H\times 10^{6} $& \multirow{2}{*}{$\Lambda$}& \multirow{2}{*}{$T_c$}&\multirow{2}{*}{$\phi_c/T_c$}&\multirow{2}{*}{$T_\text{nucl}$}& \multirow{2}{*}{$\gamma_{_w}^\text{T}$} & \multirow{2}{*}{$\alpha$} & \multirow{2}{*}{$\beta/H_*$} & \multirow{2}{*}{$H_* \tau_{\text{sh}}$}\\
         (EW scale )&($\sim 1$ TeV) & & & & & & & &\\
         \hline
         \hline
         \multicolumn{10}{|c|}{No barrier}\\
         \hline
         \hline
         $-2.526$ & $-2.537$ & $627.88$ & 110 & 1.66 & 104.5& $<1^{\textcolor{red}{*}}$  & 0.008&  4222 & 0.05 \\
         $-3.115$ & $-2.965$ & $580.73$ & 96.4 & 2.22 & 78.5 & $<1^{\textcolor{red}{*}}$  &0.029 &  780 & 0.08   \\
         $-3.375$ & $-3.137$ & $564.57$ & 90.1 & 2.49 & 52& 
         5.89  & 0.15 & 96 & 0.23\\
         $-3.389$ & $-3.147$ & $563.74$ & 89.7 & 2.50 & 46.7& 9.63  & 0.24 & 38 & 0.41\\
         $-3.392$ & $-3.148$ & $563.57$ &89.6& 2.51 & 44.2  & 11.99  & 0.30 & 15 & 0.86\\
         \hline\hline
        \multicolumn{10}{|c|}{Barrier too high}\\
         \hline\hline
    \end{tabular}
    \caption{Variation in the critical temperature ($T_c$), nucleation temperature ($T_\text{nucl}$), terminal boost factor ($\gamma_{_w}^\text{T}$), change in enthalpy ($\alpha$), speed of the phase transition ($\beta/H_*$) and sound wave lifetime ($H_* \tau_{\text{sh}}$) with respect to $C_H$, taking into account dimension 6 contribution only. The temperatures and the UV scale ($\Lambda$) have been reported in units of GeV, whereas $C_H$ is in units of $\text{GeV}^{-2}$. For each case, we have set $C_{H\square}$ = $1.56\times 10^{-7}$ GeV${}^{-2}$ and $C_{H\mathcal{D}}=-1.15\times10^{-8}$ GeV${}^{-2}$. We report the $C_H$ values both at the electroweak scale and at 1 TeV, assuming that the new physics occurs in the latter case.
    \textcolor{red}{*} $\gamma_{_w}^\text{T} < 1$ is unphysical. Such a result simply highlights the invalidity of the large velocity limit for the corresponding benchmark point.} 
    \label{table:benchmark-points}
\end{table*}

The velocity of the bubble wall is a crucial characteristic of an FOPT and it has a pronounced impact on a number of associated phenomena, such as the production of gravitational waves, baryogenesis etc.  
The expansion of the bubble is primarily dictated by (1) the difference in energies corresponding to the two minima of the effective potential and (2) the friction between the plasma particles and the bubble wall.

As the energy released due to the separation of the minima drives the bubble to expand, the pressure from the friction rises along with the increasing velocity of the wall, leading to a scenario where a terminal velocity can be attained. By balancing these two opposing forces, one can obtain a very straightforward relationship to determine the terminal velocity,
\begin{eqnarray}\label{eq:force-balance}
    \Delta V_\text{eff} &=& V^\text{false}_{T=0}(\phi) - V^\text{true}_{T=0}(\phi)\;, \nonumber\\ &=&  P (T_\text{nucl},\,v_{_w}^{\text{T}})\;.
\end{eqnarray}

The primary contribution to the friction force comes from the changes in pressure on account of the incidence of particles, traversing from the symmetric phase to the broken phase and vice versa, on the wall. The change in pressure can be expressed as \cite{BarrosoMancha:2020fay,Dine:1992wr,Azatov:2020ufh}
\begin{equation}\label{eq:pressure-change}
\Delta\mathcal{P} = \int \frac{d^3 p}{(2\pi)^3}\,n_if_i(p)\left[p_z^\text{symm.} - p_z^\text{broken}\right]|v_w|\;.
\end{equation}
Here, $n_i$ denotes the number of degrees of freedom for the $i$th particle, see Eq.~\eqref{eq:degrees-of-freedom}  and
\begin{equation}
    f_i(p) = \cfrac{1}{ e^{\gamma_{_w}\left(E + v_w\,p_z\right)/T} \pm 1 } \equiv g(E + v_w\,p_z)\;,
\end{equation}
is the distribution function with $\gamma_{_w} = \left(1-v_{_w}^2\right)^{-1/2}$. The $+$ and $-$ signs in the denominator correspond to fermions and bosons respectively. The changes in pressure, owing to particle translation, receive three distinct contributions.  
\begin{enumerate}
    \item Particles with extremely low momenta incident from the symmetric phase are reflected, consequently generating pressure on the wall. 
    \item Particles entering into the broken phase with high enough momenta end up accumulating masses. This ends up in a momentum transfer or shift in the momentum, resulting in pressure on the bubble wall. 
   \item Massive particles from the broken phase can escape through the wall into the symmetric phase and contribute to the pressure through the change in momenta.
\end{enumerate}

\begin{table*}[!ht]
    \centering
    \setlength{\tabcolsep}{12pt}
    \renewcommand{\arraystretch}{1.9}
    \begin{tabular}{|c|c|c|c|c|c|c|c|c|c|}
         \hline
         \hline
         $C_H\times 10^{6} $&$C_H\times 10^{6} $& \multirow{2}{*}{$\Lambda$}& \multirow{2}{*}{$T_c$}&\multirow{2}{*}{$\phi_c/T_c$}&\multirow{2}{*}{$T_\text{nucl}$}& \multirow{2}{*}{$\gamma_{_w}^\text{T}$} & \multirow{2}{*}{$\alpha$} & \multirow{2}{*}{$\beta/H_*$} & \multirow{2}{*}{$H_* \tau_{\text{sh}}$}\\
         (EW scale )&($\sim 1$ TeV) & & & & & & & &\\
         \hline
         \hline
         \multicolumn{10}{|c|}{No barrier}\\
         \hline
         \hline
         $-1.911$ & $-1.992$ & $708.58$ & 116.8 & 1.36 & 114.3&
         $<1^{\textcolor{red}{*}}$ & 0.005 & 10937 & 0.03 \\
         $-2.723$ & $-2.581$ & $622.40$ & 89.8 & 2.49 & 58.1&  3.04 & 0.094 & 208 & 0.15\\
         $-2.750$ & $-2.600$ & $620.23$ & 88.7 & 2.54 & 51.0& 6.94 & 0.16 & 104 & 0.20\\
         $-2.758$ & $-2.604$ & $619.69$ & 88.4 & 2.55 & 47.2& 9.96 & 0.22 & 59 & 0.28\\
         $-2.766$ & $-2.608$ & $619.16$ & 88.1 & 2.57 & 43.4&  13.86 & 0.31 & 20 & 0.66\\
         \hline\hline
        \multicolumn{10}{|c|}{Barrier too high}\\
         \hline\hline
    \end{tabular}
    \caption{Variation in the critical temperature ($T_c$), nucleation temperature ($T_\text{nucl}$) terminal boost factor ($\gamma_{_w}^\text{T}$), change in enthalpy ($\alpha$), speed of the phase transition ($\beta/H_*$), and sound wave lifetime ($H_* \tau_{\text{sh}}$) with respect to $C_H$ considering dimension 8 contributions along with dimension 6. The temperatures and the UV scale ($\Lambda$) are reported in units of GeV, whereas $C_H$ is in units of $\text{GeV}^{-2}$. For each case, we have set $C_{H\square}$ = $1.56\times 10^{-7}$ GeV${}^{-2}$ and $C_{H\mathcal{D}}=-1.15\times10^{-8}$ GeV${}^{-2}$. We report the $C_H$ values both at the electroweak scale and at 1 TeV, assuming that the new physics occurs in the latter case. It must be emphasized once again that $\phi_c/T_c > 1$ is not a robust criterion for an FOPT to occur, since  owing to large barrier height nucleation fails around $\phi_c/T_c \sim 2.6$. \textcolor{red}{*} $\gamma_{_w}^\text{T} < 1$ is unphysical. Such a result simply highlights the invalidity of the large velocity limit for the corresponding benchmark point.}
    \label{table:benchmark-points-with-dim8}
\end{table*}

In our analysis, we have investigated the possibility of generating bubbles with ultrarelativistic velocities within the framework of the SMEFT. In this limit, nearly all the particles incident from the symmetric phase have significant momenta to traverse through the wall. Hence, the dominant contribution to the pressure is generated through the inward transmission of these particles. It has been shown in Ref.~\cite{Chun:2023ezg} that the contributions from the reflection of particles on the wall as well as the from outward transmission of massive particles become negligibly small in the case of nearly relativistic bubble wall velocities.

This simplification leads to a straightforward expression for deriving the pressure in the leading order (LO) of the couplings
\begin{eqnarray}\label{eq:PLO-large-v}
    P_\text{LO}(T) &\approx&  \sum_i  n_i c_i  m^2_i(v) \,\cfrac{T^2}{24}\;.
\end{eqnarray}
Here, $m^2_i(v)$ is the mass of the $i$ th particle in the broken phase and $c_i = 1, 1/2$ for bosons and fermions respectively. We note from Eq.~\eqref{eq:PLO-large-v} that $P_{\text{LO}}$ is independent of the wall velocity $v_{_w}$. As a result, bubbles can undergo permanent accelerating (runaway) behavior~\footnote{ Reference.~\cite{Ai:2024shx} recently pointed out that the maximum friction can be larger than the leading order pressure given in Eq.~\eqref{eq:PLO-large-v} owing to the non-monotonic relation between the pressure and the velocity of the bubble. The possibility of additional pressure has also been noted in Refs.~\cite{Cline:2021iff,Laurent:2022jrs,Ai:2023see} }. 

However, this is contrary to what occurs in most realistic scenarios. An additional contribution, to the wall pressure, comes from the emission of multiple gauge bosons in frameworks where they receive masses due to an FOPT. To elaborate, this effect primarily arises from $1 \to 2$ processes that ultimately produce massive bosons within the bubble from a massless particle incident on the wall. Taking into consideration the change in the vertex factors due to $\mathcal{O}_{H\mathcal{D}}$ and $\mathcal{O}_{H\square}$, relevant for the processes involving $h$ splitting to produce massive bosons ($hhZ$ and $hhW$), the expression for NLO pressure reduces to
\begin{eqnarray}\label{eq:PNLO-large-v}
    P_\text{NLO}(T) \approx \cfrac{\kappa\,\zeta(3)}{\pi^3}\,\gamma_w \, m_{_Z}(v_h)\, T^3\,\log\left(\cfrac{m_Z(v)}{g_2 T}\right)\;,
\end{eqnarray}
where
\begin{equation} \label{eq:kappa}
\kappa \sim (5+0.4v^2\,C_{H\square}-0.2v^2\,C_{H\mathcal{D}})\;.
\end{equation}
For multiple gauge boson emissions, a careful resummation is required from the real and virtual processes~\cite{Gouttenoire:2021kjv}. It is evident from Eq.~\eqref{eq:PNLO-large-v} that $P_{\text{NLO}}$ starts to become substantial when a certain velocity threshold is surpassed. Thus, according to  Eq.~\eqref{eq:force-balance}, the bubbles reach a terminal velocity when the combined friction effect from Eqs.~\eqref{eq:PLO-large-v} and~\eqref{eq:PNLO-large-v} becomes large enough to overcome $\Delta V_{\text{eff}}$,
 \begin{eqnarray}
     P_{\text{NLO}}(\gamma_{_w}^{\text{T}},T_\text{nucl}) = \Delta V_{\text{eff}} - P_{\text{LO}}(T_\text{nucl})\;.
 \end{eqnarray}

Throughout our calculation we use fixed values for $C_{H\square}$ and $C_{H\mathcal{D}}$, i.e., $C_{H\square}$ = $1.56\times10^{-7}$ GeV$^{-2}$ and $C_{H\mathcal{D}}$ = $-1.15\times10^{-8}$ GeV$^{-2}$. Incorporating these values in Eq.~\eqref{eq:PNLO-large-v} we find a simplified formula for estimating the boost factor for bubbles in an ultrarelativistic scenario, 
 \begin{align}
     \gamma_{w}^{\text{T}} \approx 2\times&\bigg[\cfrac{(\Delta V_{\text{eff}} - P_{\text{LO}})}{m_W^4}\bigg]\bigg[\cfrac{100\,\text{GeV}}{T_\text{nucl}}\bigg]^3\left[\log\bigg(\cfrac{m_{_Z}}{g_2T_\text{nucl}}\bigg)\right]^{-1}\;.
 \end{align}

For both sets of benchmark points, i.e., those listed in  Table~\ref{table:benchmark-points}, and Table~\ref{table:benchmark-points-with-dim8}, we highlight the cases where such ultrarelativistic bubble wall velocities can be obtained. We notice that this happens for a very narrow region of the parameter space. This is instructive for the possibility of simultaneously testing relevant SMEFT operators at multiple frontiers, including the energy and cosmic frontiers.

The impact of RG evolution was once again tested by running the parameters down from $\mu = v$ to $\mu = m_t$. We noticed that this led to shifts in the values of $C_H$, for the last 3 entries of Table~\ref{table:benchmark-points}, into the high barrier region highlighted on the left side of Fig.~\ref{fig:chbox_vs_ch}, therefore, disallowing not only the production of ultrarelativistic bubbles but even an FOPT for such parameter choices.

It must be stressed that one must check if the phase transition is complete before the disappearance of the potential barrier. In other words, the time of phase transition can be approximated in terms of the radius of the bubble at the time of percolation, which is approximately given by~\cite{Enqvist:1991xw,Azatov:2019png}
\begin{eqnarray}
    R_\text{per} \simeq \frac{(8\pi)^{1/3}}{\widetilde{\beta}_\text{per}}\;, \; \; \text{where}\;\; \; \widetilde\beta_\text{per} \simeq \frac{\beta}{H}\Big|_{T_\text{per}}\;,
\end{eqnarray}
where $H$ denotes the Hubble parameter at the time of percolation and $T_\text{per}$ is defined as the temperature for which roughly $\sim 30\%$ of the space has been converted to the true broken phase. $\widetilde\beta$ is the inverse time duration for the phase transition. It is defined in terms of the bubble nucleation rate $\,\,\Gamma(t)$ as, $\beta \approx \dot{\Gamma}/\Gamma$, and it can be estimated using the temperature slope of the three-dimensional bounce action, see [Eq.~\eqref{eq:action}], as
\begin{eqnarray}\label{eq:beta-per}
\frac{\beta}{H}\Big|_{T_\text{per}} = T_\text{per}\, \cfrac{d}{dT} \,\bigg(\cfrac{1}{T}\, S(T)\bigg)_{T=T_\text{per}}\;.
\end{eqnarray}
Ignoring the difference between percolation and nucleation temperature, we find $\widetilde\beta\in (10^1, 10^2)$ for the parameter region of interest. Consequently, the drop in temperature associated with the expansion of the bubble is given by~\cite{Azatov:2022tii}
\begin{eqnarray}
    \Delta T \sim T_\text{nucl} H \Delta t \sim \frac{T_\text{nucl} H}{\beta}\;.
\end{eqnarray}

We note that the sufficiently large values of $\beta/H$ for our choice of parameters ensure that the temperature drop resulting from the bubble expansion is minuscule, i.e., $O(10^{-1}) - O(1)$, and hence does not lead to the disappearance of the barrier. 


\section{Stochastic gravitational wave background}\label{sec:SGWB}

A strong FOPT produces a GWS through three distinct mechanisms~\cite{Caprini:2015zlo,Caprini:2019egz,Caprini:2007xq,Huber:2008hg,Caprini:2009fx,Katz:2016adq}. 
\begin{enumerate}
    \item The collision of bubble walls can lead to a substantial contribution to the GWS. Although, this depends on whether the bubbles continue to accelerate until the time of collision.

    \item A second contribution arises from the propagation of sound waves in the plasma. Based on simulations, it has been noted that sound waves usually produce a GWS for longer periods of time than bubble wall collisions.

    \item If the duration of propagation of the sound waves is not that extensive, then magnetohydrodynamic (MHD) turbulence in the plasma also acts as a source for the GWS.
\end{enumerate}

The overall stochastic GW background signal owing to a first order phase transition can then be given as

\begin{eqnarray}\label{eq:GW-background}
h^2 \Omega_{\text{GW}} = h^2 \Omega_{\text{col}} + h^2 \Omega_{\text{sw}} + h^2 \Omega_{\text{turb}}\;,
\end{eqnarray}

where the individual terms on the right side can be quantified using the following formulas \cite{Caprini:2007xq,Huber:2008hg,Caprini:2009fx,Katz:2016adq}:
\begin{widetext}
\begin{eqnarray}\label{eq:GWS-contributions}
h^2 \Omega_{\text{col}}(f)  &=& 1.67 \times 10^{-5} \, \times 
	\bigg(\cfrac{\beta}{H_*}\bigg)^{-2} \bigg(\cfrac{\kappa_\text{col}\, \alpha}{1 + \alpha}\bigg)^2  \bigg(\cfrac{g_*}{100}\bigg)^{-\frac{1}{3}} \bigg(\cfrac{0.11 v_w^3}{0.42 + v_w^2}\bigg) \bigg[\cfrac{3.8\,(f/f_{\text{col}})^{2.8}}{1 + 2.8\,(f/f_{\text{col}})^{3.8}}\bigg]\;,\nonumber\\
    h^2 \Omega_{\text{sw}}(f)  &=& 2.65 \times 10^{-6} \times v_w \, \bigg(\cfrac{\beta}{H_*}\bigg)^{-2}  \bigg(\cfrac{\kappa_\text{sw} \,\alpha}{1 + \alpha}\bigg)^2  \bigg(\cfrac{g_*}{100}\bigg)^{-\frac{1}{3}}   \bigg[\cfrac{f}{f_{\text{sw}}}\bigg]^3 \bigg[\frac{7}{4 + 3 (f/f_{\text{sw}})^2}\bigg]^{7/2}\;, \\
    h^2 \Omega_{\text{turb}}(f)  &=& 3.35 \times 10^{-4}  \times  v_w \, \bigg(\cfrac{\beta}{H_*}\bigg)^{-1}  \bigg(\cfrac{\kappa_\text{turb}\,\alpha}{1 + \alpha}\bigg)^{3/2}  \bigg(\cfrac{g_*}{100}\bigg)^{-\frac{1}{3}} \bigg[\cfrac{(f/f_{\text{turb}})^3}{[1 + (f/f_{\text{turb}})]^{11/3}\,(1 + 8 \pi f/h_*)}\bigg]\;.\nonumber
\end{eqnarray}
\end{widetext}
In each of the above equations, $\beta$ is now defined at the nucleation temperature, i.e.,
\begin{eqnarray}\label{eq:beta}
\cfrac{\beta}{H_*} = T_\text{nucl}\, \cfrac{d}{dT} \,\bigg(\cfrac{1}{T}\, S(T)\bigg)_{T=T_\text{nucl}}\;,
\end{eqnarray}
and $H_*$ is the Hubble parameter when the gravitational waves are produced. 

\begin{figure*}[!t]
    \centering
    \includegraphics[width=\linewidth]{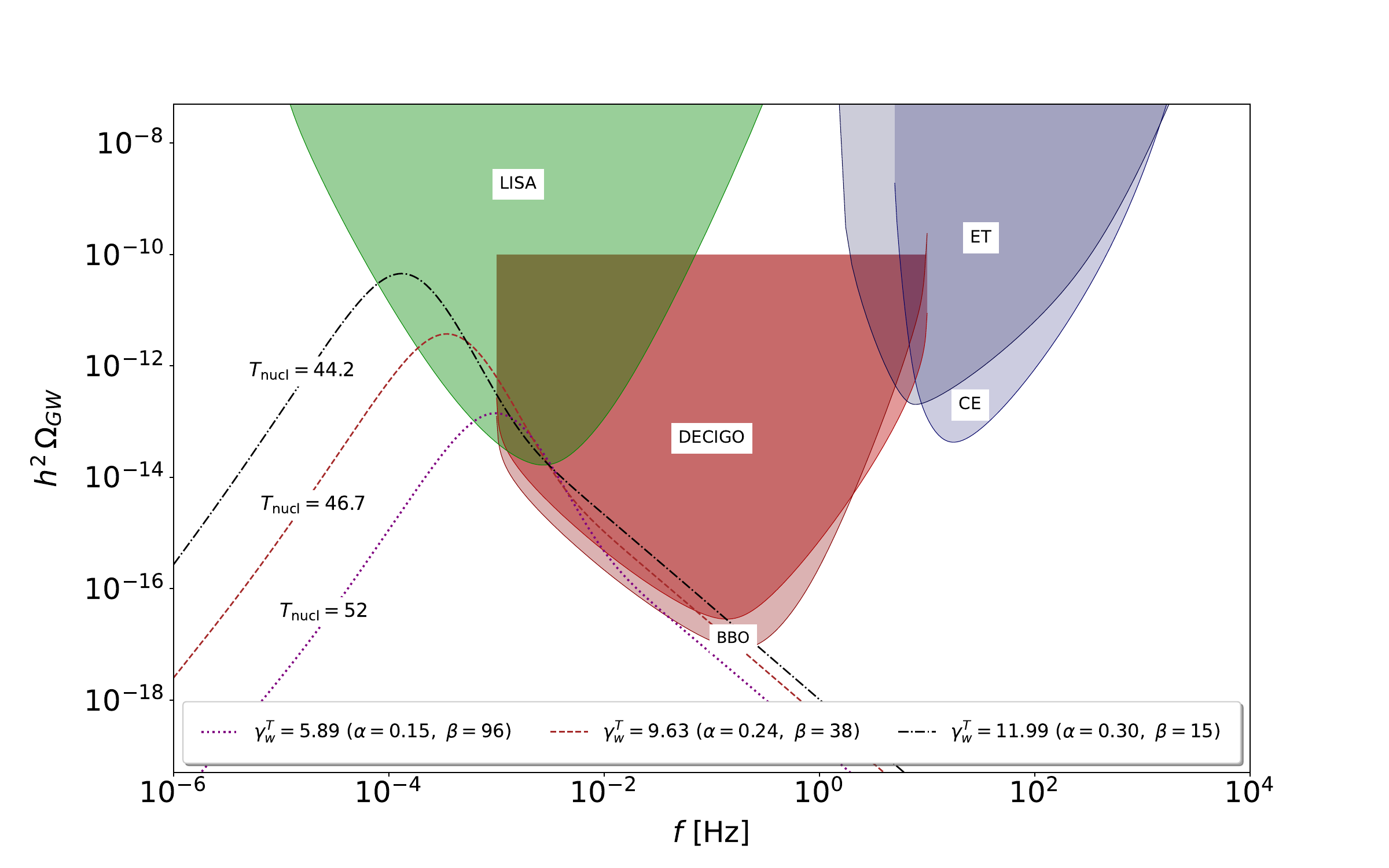}
    \caption{Stochastic gravitational wave spectra as a function of frequency for different benchmark points, plotted against the projected power law integrated sensitivities of LISA \cite{LISA:2017pwj,Caprini:2015zlo,Caprini:2019egz}, DECIGO \cite{Seto:2001qf}, BBO \cite{Corbin:2005ny,Crowder:2005nr}, Einstein Telescope \cite{Punturo:2010zz} and Cosmic Explorer \cite{Evans:2021gyd} experiments. The benchmark points corresponding to the case with a dimension 8 term in the Lagrangian, yield similar values for ($T_\text{nucl}$, $\gamma_w^\text{T}$, $\alpha$ and $\beta/H_*$). Therefore, the gravitational wave spectra do not present any noticeable changes.}
    \label{fig:gws}
\end{figure*}

 On the other hand, $\alpha$ defines the change in enthalpy associated with the FOPT, and it is given as the ratio
between the vacuum energy and the total energy stored in radiation
\begin{eqnarray}\label{eq:alpha}
\alpha &=& \frac{1}{\rho_\text{rad}}\,\left[\Delta\,V(T) - \frac{T}{4}\,\frac{\partial\, \Delta\, V(T)}{\partial\, T}\right]\Bigg|_{T = T_\text{nucl}}\;,
\end{eqnarray}  
with $\rho_\text{rad} \sim 0.03 \times T_\text{nucl}^{-4}$
and $\Delta V(T) = V_\text{false} - V_\text{true}$ being the depth of the true minima of the effective thermal potential. The $\alpha$ and $\beta/H_*$ values corresponding to each benchmark point have also been catalogued in Tables~\ref{table:benchmark-points} and \ref{table:benchmark-points-with-dim8}. $\kappa_\text{col}$, $\kappa_\text{sw}$, and $\kappa_\text{turb}$ are efficiency factors corresponding to the bubbles, the bulk fluid, and the turbulence in the plasma. These can be expressed as functions of $\alpha$ based on whether we are dealing with \textit{runaway} or \textit{nonrunaway} bubbles. 

To ascertain the \textit{runaway} versus \textit{nonrunaway} nature of the bubbles, 
we estimate the Lorentz factor in the absence of any NLO friction, $\gamma_\star^w$, just before the collision. Note that, $\gamma_\star^w$ controls the surface energy of the bubbles. By equating the surface energy to the gain in the potential energy, one finds:
\begin{eqnarray}
    \gamma_{\star}^w \simeq \frac{2R_\star}{3R_0} \left(1-\frac{P_{LO}}{\Delta V}\right)\;.
    \label{eq:gamma-star}
\end{eqnarray}
Here, $R_0\sim 1/T_\text{nucl}$ is defined as the bubble size during nucleation. On the other hand, $R_\star$ is the bubble size at the time of the collision, and under the assumption that $v_w\to 1$ is given by
\begin{eqnarray}
    R_{\star} \sim  \frac{1}{\beta} \sim \frac{1}{\widetilde{\beta} H} \sim O(10^{-1}-10^{-2}) H^{-1}\;,
\end{eqnarray}
with $\beta$ defined in Eq.~\eqref{eq:beta}.

Depending on the relative strengths of $\gamma^w_\star$ and $\gamma_w^{\text{T}}$, different contributions to the GWS become important. For example, if $\gamma^w_\star>\gamma_w^{\text{T}}$, bubbles reach the equilibrium or terminal boost factor before the collision and the contribution from the wall collision to the GW spectrum becomes minuscule. This corresponds to the \textit{nonrunaway} case for which
\begin{eqnarray}\label{eq:GW-background-nonrunaway}
h^2 \Omega_{\text{GW}} \sim h^2 \Omega_{\text{sw}} + h^2 \Omega_{\text{turb}}\;.
\end{eqnarray}
On the other hand, $\gamma_w^{\text{T}}>\gamma^w_\star$ signifies that the collision of the bubbles happens even before the equilibrium terminal boost can be achieved. In such a case, the GW spectrum receives a dominant contribution from bubble collisions as well. 

For each of the benchmark points of interest, i.e., the last three entries of Table~\ref{table:benchmark-points}, we find that $\gamma^w_\star \gg \gamma_w^{\text{T}}$. Therefore, each of these corresponds to the case of \textit{nonrunaway} bubbles and the efficiency factors for the three sources of the GWS can be expressed as
\begin{eqnarray} 
\kappa_\text{col} &\rightarrow& 0\;, \nonumber\\
\kappa_\text{sw} &=& \alpha \, (0.73 + 0.083 \times \sqrt{\alpha} + \alpha)^{-1}\;, \nonumber\\
\kappa_\text{turb} &=& \epsilon\,\kappa_\text{sw}\;,
\end{eqnarray}

with $\epsilon \sim 0.05-0.1$ representing that small fraction of the bulk motion which is turbulent. 

It has been pointed out that the sound wave lifetime is notably smaller for weaker phase transitions, i.e. those characterized by small $\alpha$ values. This affects the sound wave contribution, $h^2 \Omega_{\text{sw}}(f)$, in Eq.~\eqref{eq:GWS-contributions}, which in such cases is proportional to $\beta^{-2}$. The sound wave lifetime can be estimated as \cite{Ellis:2020awk},
\begin{eqnarray}
    H_* \tau_{\text{sh}} = \cfrac{(8\pi)^{1/3}\times\text{Max}(v_w, c_s)}{v_\text{rms}}\,\bigg(\cfrac{\beta}{H_*}\bigg)^{-1}\;,
\end{eqnarray}
where $c_s$ is the speed of sound in plasma ($1/\sqrt{3}$ for relativistic fluid) and the root-mean-square fluid velocity $v_{\text{rms}}$ is given in terms of $\kappa_\text{sw}$ and $\alpha$ as
\begin{eqnarray}
  v_{\text{rms}} \simeq \sqrt{\cfrac{3}{4}\,\cfrac{\kappa_\text{sw}\,\alpha}{1+\alpha}}\;.
\end{eqnarray}
We have added the values of $H_* \tau_{\text{sh}}$ corresponding to all benchmark points in Tables~\ref{table:benchmark-points} and \ref{table:benchmark-points-with-dim8} and we note that $H_* \tau_{\text{sh}} < 1$ for each case, which indicates a short lifetime for the sound waves \cite{Caprini:2019egz}.

Lastly, the peak frequencies corresponding to each contribution in Eq.~\eqref{eq:GWS-contributions} are given as,
\begin{eqnarray}\label{eq:GWS-peak-f}
f_{\text{col}}  &=& 16.5 \times 10^{-6}\,  \bigg(\cfrac{0.62}{1.8 - 0.1 v_w + v_w ^2}\bigg) \, g(T_\text{nucl}) \, \text{Hz}\;, \nonumber\\
f_{\text{sw}} &=& 1.9 \times 10^{-5}\, \bigg(\cfrac{1}{v_w}\bigg) \, g(T_\text{nucl}) \,\text{Hz}\;, \nonumber\\
f_{\text{turb}} &=& 2.7\times 10^{-5}\, \bigg(\cfrac{1}{v_w}\bigg) \, g(T_\text{nucl}) \,\text{Hz}\;.
\end{eqnarray}
with
\begin{eqnarray}
    g(T) = \bigg(\cfrac{\beta}{H_*}\bigg) \bigg(\cfrac{T}{100 \text{ GeV}}\bigg) \bigg(\cfrac{g_*}{100}\bigg)^{1/6}\;,
\end{eqnarray}
and $h_{*}$ in Eq.~\eqref{eq:GWS-contributions} represents the redshifted Hubble time
\begin{eqnarray}
    h_{*}  &=& 16.5 \times 10^{-6}\bigg(\cfrac{T_\text{nucl}}{100 \text{ GeV}}\bigg)\, \bigg(\cfrac{g_*}{100}\bigg)^{1/6}\text{Hz}\;.
\end{eqnarray}
Using these ingredients, we have estimated the stochastic gravitational wave spectrum for the relevant benchmark points and we have plotted those against the projected sensitivities of upcoming gravitational wave detection experiments in Fig.~\ref{fig:gws}. For each curve, we have also highlighted the $\alpha$ and $\beta$ values.

\section{Conclusion}\label{sec:conclusion}
 Recent studies have shown that ultrarelativistic bubble expansion during first order phase transition can play a very important role in addressing baryogenesis and dark matter production. Such bubbles also leave telltale signatures in the gravitational wave spectrum. It is well-known that many scalar extensions of the SM with a two-step phase transition can accommodate such bubble velocities. However, the absence of any direct evidence for physics beyond the SM has prompted us to investigate the prospect of ultrarelativistic bubbles in the framework of the SMEFT. In this work, we augment the SM with mass dimension 6 operators constructed solely of the Higgs and its derivatives such as $\mathcal O_H$, $\mathcal{O}_{H\mathcal{D}}$, and $\mathcal{O}_{H\square}$. Out of these, $\mathcal{O}_H$ plays a crucial in FOPT as it contributes to the Higgs potential directly, whereas the effect of other operators comes through wave function renormalization of the Higgs field. We studied the impact of such operators in the renormalization group improved one-loop effective potential in addition to temperature-dependent corrections. As expected, $\mathcal{O}_H$ has the most pronounced impact as far as the thermal potential and nucleation temperature are concerned. It is to be noted that a lower nucleation temperature, in comparison to the electroweak scale, is preferred from the perspective of ultrarelativistic bubbles. We found that such a possibility can only be accommodated in a narrow region of $C_H$. The primary reason can be attributed to the fact that for lower values of $C_H$, the potential barrier becomes too high. Whereas, in the opposite limit, the barrier disappears. This is interesting as the usual global fit constraints on $C_H$ are rather weak and the strongest constraints can be obtained by assuming ultrarelativistic bubbles. In addition, such a range for $C_H$ would also leave its fingerprint in the gravitational wave spectrum. It is worth mentioning that the usual criterion of strong FOPT, i.e., $\phi_c/T_c\gtrsim1$ should be carefully analyzed when commenting on the successful completion of the phase transition. In our analysis, we found that this ratio only falls between the values $1.66$ and $2.51$ in cases where bubble nucleation is possible, which is, again, mainly controlled by the height of the barrier. It must be emphasized that there is a quantifiable impact of the RG evolution of running parameters on nucleation temperatures for different benchmark points as well as on the corresponding bubble wall velocities. We noted changes of the order of $4 - 20\%$ in the nucleation temperature when varying the renormalization scale. In fact, some previously allowed parameter choices are ruled out once the running parameters are brought down to a lower scale.
 
 Our results on the ranges of $C_H$ can also be translated to the new physics scale $\Lambda$, which in turn helps us to understand the specifics of the UV completions that could lead to distinct characteristics of FOPT. The requirement of a successful phase transition sets the scale $\Lambda$ to be in the range of $563.57$-$627.88$ GeV for the $O(1)$ $C_H$ coefficient, We also found that this gets shifted to the slightly higher range of $619.16$ GeV $\leq \Lambda \leq$ 708.58 GeV when taking into account a dimension 8 term in the Lagrangian as well. We notice that, within these ranges, the phase transition can actually lead to ultrarelativistic bubbles only when $\Lambda$ falls between a small interval of $563.57$-$564.57$ GeV for the dimension 6 only case and within $619.16$-$622.40$ GeV for the case with a dimension 8 term. Such fine-tuned values obviously require proper UV completion, and this can be achieved with scalar extensions of the SM. For example, the real singlet scalar extension would generate $C_H\sim A^2 \kappa/M^4$ and $C_{H\Box}\sim A^2/M^4$, where $A$ and $\kappa$ are the couplings of $|H|^2\phi$ and $|H|^2\phi^2$, respectively, and $M$ is the mass of the additional scalar. Therefore, using these matching relations, the constraints for successful bubble nucleation can be easily translated onto the UV model parameters. It is also noteworthy that the requirement for bubble nucleation poses a more sensitive bound on the parameter spaces of those UV models that produce at least $\mathcal{O}_H$ at tree-level, compared to those that generate it at one-loop (e.g., SM extension with a complex singlet scalar). However, it must be noted that although the main characteristics of a phase transition, informed by different UV models, can be encoded within an EFT, one must be cautious while establishing one-to-one correspondences between the predictions of the full theory versus those of the corresponding EFT. 

\acknowledgments
S.C. would like to thank Science and Engineering Research Board, Government of India (Grant No.: SRG/2023/001162) for financial support. S.P. acknowledges the support by the MHRD, Government of India, under the Prime Minister's Research Fellows (PMRF) Scheme 2020, and by the Spanish Ministerio de Ciencia, Innovación y Universidades through Grants No. PID2020-114473GBI00 and No. CNS2022-135595. U.B. is grateful to the Mainz Institute for Theoretical Physics (MITP) of the Cluster of Excellence $\text{PRISMA}^+$ (Project ID 39083149) for its hospitality and support. The authors would like to thank Carlos Tamarit for his input regarding the code for the shooting method. We would also like to thank Joydeep Chakrabortty and Miguel Vanvlasselaer for helpful discussions. We also thank Aleksandr Azatov for comments and suggestions on the draft.

\appendix
\section{Comments on $ P_\text{NLO}(T)$ calculation:}

We have closely followed the calculation for $P_\text{NLO}(T)$ as outlined in \cite{Azatov:2022tii,Gouttenoire:2021kjv}. By comparing Eq.~(28) of \cite{Azatov:2022tii} and Eq.~\eqref{eq:kappa} in this paper, we can ascertain that in the absence of EFT operators (i.e. with only SM contributions), $\kappa$ will have the following form,
\begin{eqnarray}\label{eq:app_kappa}
    \kappa \sim & 8 \,\alpha \Bigg( \cfrac{7 + 14\,c_w}{s_w^2} - \cfrac{7-15\,s_w^2}{c_w^2} \Bigg),
\end{eqnarray}
where, $s_w (c_w) \equiv \sin{\theta_w} (\cos{\theta_w})$, $\theta_w$ is the weak mixing angle. If we take $\cos{\theta_w} = m_W/m_Z$, $\kappa \to 5$, which is the value quoted in Eq.~\eqref{eq:kappa} without the SMEFT Wilson Coefficients. The factor within the parentheses in Eq.~\eqref{eq:app_kappa} comes from SM interactions which get modified due to the effective operators. As previously mentioned, the relevant processes involving $h$ splitting are $hhZ, \text{ and } hhW$. Incorporation of the SMEFT operators introduces the field redefinitions, $h \to Z_h^{-1}\,h $ and $ G^0 \to Z_G^{-1} G^0$, \cite{Camargo-Molina:2021zgz,Dedes:2017zog} where,
\begin{eqnarray}
    Z_h &=& (1-C_{H\square}\,v^2+\cfrac{1}{4}\,C_{H\mathcal{D}}\,v^2)\;, \nonumber \\
    Z_G &=& (1+\frac{1}{4}\,C_{H\mathcal{D}}\,v^2)\;.
\end{eqnarray}
This affects the following vertices, $ G^0 G^{\pm} W^{\mp}$, $h G^{\pm} W^{\mp}$, $G^0hZ$ and consequently the expression in the parentheses in Eq.~\eqref{eq:app_kappa} gets modified. It must be noted however that another relevant vertex $G^+G^-Z$ does not get any EFT contribution. The process $hhW$ for which the SM contribution was $2c_w/s_w^2 $ changes to  $ c_w (Z_h^{-2}+Z_G^{-2})/s_w^2 \sim 2\,c_w (1+C_{H\square}\,v^2 -0.5\,C_{H\mathcal{D}}\,v^2 )/s_w^2.$ Similarly, for the $hhZ$ process, the $hG^0Z$ vertex is modified by $Z_h^{-1}\,Z_G^{-1}$ factor. After the suitable modifications to Eq.~\eqref{eq:app_kappa}, we obtain the coefficient of $C_{H\mathcal{D}}\,v^2$ as $-4\,\alpha(c_w\,s_w^{-2} + 0.5\,c_w^{-2}\,s_w^{-2})$ and the coefficient of $C_{H\square}\,v^2$ as $4\,\alpha(2\,c_w\,s_w^{-2} + c_w^{-2}\,s_w^{-2})$. Their numerical values are shown in Eq.~\eqref{eq:kappa}.

\addcontentsline*{toc}{section}{}
\bibliography{references}

\end{document}